\documentclass[usenatbib,letterpaper]{mn2e}
\usepackage[totalwidth=480pt,totalheight=680pt,twoside=false,voffset=.5cm]{geometry}

\usepackage{graphics,graphicx,epsfig,ulem,multirow,appendix,lscape,amsmath,amssymb,color}

\begin{document}

\title[Eccentric planet-debris disc interaction]
  {Dynamical evolution of an eccentric planet and a less massive debris disc}
\author[T. D. Pearce \& M. C. Wyatt]
  {Tim D. Pearce\thanks{tdpearce@ast.cam.ac.uk} and Mark C. Wyatt\\
  Institute of Astronomy, University of Cambridge, Madingley Road, Cambridge, CB3 0HA, UK}
\date{Released 2002 Xxxxx XX}

\pagerange{\pageref{firstpage}--\pageref{lastpage}} \pubyear{2002}

\def\LaTeX{L\kern-.36em\raise.3ex\hbox{a}\kern-.15em
    T\kern-.1667em\lower.7ex\hbox{E}\kern-.125emX}

\newtheorem{theorem}{Theorem}[section]

\interfootnotelinepenalty=10000

\label{firstpage}

\maketitle

%%%%%%%%%%%%%%%%%%%%%%%%%%%%%%%%%%%%%%%%%%%%%%%%%%%%%%%%%%%%%%%%%%%%%%%%%%%%%%%%%%%%%
%%%%%%%%%%%%%%%%%%%%%%%%%%%%%%%%%%%%%%%%%%%%%%%%%%%%%%%%%%%%%%%%%%%%%%%%%%%%%%%%%%%%%

\begin{abstract}      
        
\noindent We investigate the interaction between an eccentric planet and a less massive external debris disc. This scenario could occur after planet-planet scattering or merging events. We characterise the evolution over a wide range of initial conditions, using a suite of n-body integrations combined with theory. Planets near the disc mid-plane remove the inner debris region, and surviving particles form an eccentric disc apsidally aligned with the planet. The inner disc edge is elliptical and lies just beyond the planet's orbit. Moderately inclined planets ($i_{\rm plt} \gtrsim 20^\circ$ for $e_{\rm plt} = 0.8$) may instead sculpt debris into a bell-shaped structure enveloping the planet's orbit. Finally some highly inclined planets ($i_{\rm plt} \sim 90^\circ$) can maintain a disc orthogonal to the planet's plane. In all cases disc particles undergo rapid evolution, whilst the overall structures evolve more slowly. The shapes of these structures and their density profiles are characterised. The width of the chaotic zone around the planet's orbit is derived in the coplanar case using eccentric Hill radius arguments. This zone is cleared within approximately ten secular or diffusion times (whichever is longer), and debris assumes its final shape within a few secular times. We quantify the planet's migration and show it will almost always be small in this mass regime. Our results may be used to characterise unseen eccentric planets using observed debris features.

\end{abstract}

\begin{keywords}
planet-disc interactions - planets and satellites: dynamical evolution and stability - circumstellar matter
\end{keywords}

%%%%%%%%%%%%%%%%%%%%%%%%%%%%%%%%%%%%%%%%%%%%%%%%%%%%%%%%%%%%%%%%%%%%%%%%%%%%%%%%%%%%%
%%%%%%%%%%%%%%%%%%%%%%%%%%%%%%%%%%%%%%%%%%%%%%%%%%%%%%%%%%%%%%%%%%%%%%%%%%%%%%%%%%%%%

\section{Introduction}
\label{sec: introduction}

\noindent Extrasolar planets exhibit a broad range of eccentricities, including values far higher than those typical of present day Solar System planets\footnote{See the exoplanet.eu database}. However planets should form on circular orbits as eccentricity excitations are damped by the protoplanetary disc \citep{Lissauer93}, so high eccentricities are generally attributed to post-formation dynamical processes \citep{Tremaine04}. Solar System dynamical theories suggest that the giant planets underwent significant migration and orbital evolution at early times \citep{Tsiganis05, Walsh11}, and such models could apply to other star systems too. During such violent periods eccentricity driving interactions, such as planet-planet scattering \citep{Ford08} and mergers \citep{Lin97}, could be common-place. Stellar flybys \citep{Malmberg09}, planet-disc interactions \citep{Bitsch13} and long-term interactions between planets \citep{Wu11} could also drive up planet eccentricities. So with large eccentricity changes possible, planets could come into contact with bodies previously too distant for interactions to occur.

In this paper, we investigate the scenario where such an eccentric planet encounters a debris disc. The latter describes a circumstellar disc of solids, containing bodies ranging in size from sub-micrometer dust grains all the way up to dwarf planets, and which is probably composed of material left over from the protoplanetary disc (see \citealt{Wyatt08} for a comprehensive review). The Solar System's asteroid and Kuiper belts fit this description. Debris material forms a collisional cascade, such that the collisional breakup of objects produces ever smaller bodies which are eventually blown out of the system by radiation pressure \citep{Backman93, WyattDent02}. The dust re-emits starlight at longer wavelengths, and of order a thousand extrasolar debris discs have been detected as mid- to far-infrared excesses (e.g. \citealt{Patel14}). Observed discs typically contain $0.01 - 1$ Earth masses (${\rm M}_\oplus$) in emitting grains \citep{Wyatt03}, and extrapolating these to include larger unseen bodies yields total masses several orders of magnitude greater (e.g. \citealt{WyattDent02}). The discs are observed at a few to hundreds of au from their host stars \citep{Najita05}; for context, the total mass of the early Kuiper belt is thought to have been tens of ${\rm M}_\oplus$ spread between 30 and 50 au \citep{Stern97}.

Many debris discs have been resolved and show eccentricity and/or density asymmetries, which are often attributed to perturbations by unseen planets (e.g. \citealt{Kalas05, Krist12, Schneider09}). The effect of a low eccentricity, low inclination planet on a cold debris disc is well studied, and invoking such planets successfully explains observed debris disc features in several systems such as offsets, warps and spirals (e.g. \citealt{Wyatt99, Augereau01, Wyatt05}). However discs with large asymmetries (e.g. \citealt{Eiroa10}) or substantial misalignments (e.g. \citealt{Kennedy12}) have also been observed which, combined with the discovery of Fomalhaut b (a highly eccentric and possibly disc crossing object: \citealt{Kalas13, Beust14}), motivate the study of more extreme interactions. Studies involving a highly eccentric planet have recently been conducted by \cite{Beust14}, \cite{Tamayo14} and \cite{Faramaz14}, specifically attempting to explain the discs of Fomalhaut and $\zeta^2$ Reticuli. We wish to characterise this interaction generally, to produce quantitative predictions applicable over a broad region of parameter space. This includes planets with large initial inclinations and those with non-negligible orbital evolution. In this paper we only examine planets which are more massive than the disc.

The layout of this paper is as follows. We describe our simulations in Section \ref{sec: simulations}, and outline the three primary outcomes in Section \ref{sec: results}. In Section \ref{sec: dynamical effects} we provide an overview of dynamical theory, and apply this in Section \ref{sec: analysis} to make predictions about the post-interaction debris structure, planet evolution and important timescales. We discuss our results in Section \ref{sec: discussion} and conclude in Section \ref{sec: conclusions}. We use the subscript ``plt'' to denote parameters of the planet throughout the paper, and colour versions of all plots are available in the online version.

%%%%%%%%%%%%%%%%%%%%%%%%%%%%%%%%%%%%%%%%%%%%%%%%%%%%%%%%%%%%%%%%%%%%%%%%%%%%%%%%%%%%%
%%%%%%%%%%%%%%%%%%%%%%%%%%%%%%%%%%%%%%%%%%%%%%%%%%%%%%%%%%%%%%%%%%%%%%%%%%%%%%%%%%%%%

\section{Simulations}
\label{sec: simulations}

\noindent We modelled the interaction using the Mercury 6.2 n-body integrator \citep{Chambers99}, with a hybrid symplectic / Bulirsch-Stoer algorithm. This allowed the rapid computation of distant interactions whilst maintaining accuracy in close encounters. Our systems consisted of a massive particle representing the planet and $10^3$ to $10^4$ massive test particles to model the disc. The planet therefore perturbed the disc and \textit{vice-versa} however the disc had no self-gravity, which is valid as the more massive planet would dominate over debris-debris interactions (e.g. \citealt{Beust14}). Relatively small numbers of particles let us run many integrations, and identical systems simulated with $10^3$ and $10^4$ particles resulted in the same outcome in terms of disc and planet evolution. Furthermore the planets did not evolve stochastically, so we are confident that our particle numbers are sufficiently high. We used a wide range of disc and planet masses ($M_{\rm disc}$ and $M_{\rm plt}$ respectively), with the former 0.1 to 100 ${\rm M}_\oplus$ and the latter 1 to 1000 ${\rm M}_\oplus$. The planet was at least ten times as massive as the disc. The stellar masses ($M_*$) were equal to a solar mass (${\rm M}_\odot$) in almost all integrations. Simulations typically lasted for $10^7 - 10^8$ years, long enough for the major evolution to have occurred (see Section \ref{sec: timescales}).

The discs were what we might consider early Kuiper-belt analogues, typically with inner and outer edges at $r_1 = 20$ and $r_2 = 60$ au respectively. Each disc particle was assigned an initial semi-major axis $a$ within this range. The particles themselves were of equal mass, and semi-major axes were populated in such a way that the resulting surface density profile (if each particle had negligible eccentricity $e$ and inclination $i$, the latter relative to the initial disc midplane) could be written as $\Sigma (r) \propto r^{-\gamma}$, where $r$ is the radial distance from the star and $\gamma$ is the surface density index. We used $\gamma = 1.5$, that of the Minimum Mass Solar Nebula \citep{Hayashi81}. The discs were relatively dynamically cool before the interaction, with inclinations and eccentricities typically drawn from the ranges $0-5^\circ$ and $0-0.08$ respectively. Finally, the initial argument of pericentre $\omega$, longitude of ascending node $\Omega$ and mean anomaly of each disc particle were uniformly drawn between $0$ and $360^\circ$ (note that due to the initial symmetry of the disc, the planet's ascending node was defined to lie along the $x$ axis without loss of generality).

The eccentric planets typically had pericentres at 5 au, the distance of Jupiter in our solar system and the location at which massive planets could be expected to form based on snow line arguments (e.g. \citealt{Ida04}). This is therefore a reasonable estimate for the pericentre of a body made eccentric by planet-planet interactions. Their apocentre distances were typically between the disc's initial inner and outer edge radii, with eccentricities from $0.2-0.9$. Half of our simulations had planets with orbits initially located in the disc midplane, and the other half had planets inclined by up to $90^\circ$. The latter had initial arguments of pericentre (the orientation of the planet's pericentre relative to the ascending node, where the orbit crosses the disc midplane) equal to 0, 45 or $90^\circ$. The planets were prograde with respect to the disc particles in all simulations (i.e. $0 \leq i_{\rm plt} \leq 90^\circ$).

%%%%%%%%%%%%%%%%%%%%%%%%%%%%%%%%%%%%%%%%%%%%%%%%%%%%%%%%%%%%%%%%%%%%%%%%%%%%%%%%%%%%%

\section{Results}
\label{sec: results}

\noindent We observed three outcomes in our simulations. Firstly if the planet's orbit lay roughly in the disc midplane, the inner disc particles were ejected and remaining debris formed an elliptical disc apsidally aligned with the planet (outcome Ia). Secondly if the planet's orbit brought it far out of the midplane then surviving debris instead formed a hollow bell-shaped structure, again aligned with the planet. This structure completely enveloped the planet's orbit, apart from for two holes at its pericentre and apocentre ends (outcome Ib). An extreme case of this outcome occurred for very highly inclined planets, where the holes opened to the extent that debris formed a disc orthogonal to the planet's orbit. Finally a planet could eject all non-resonant disc particles if it came sufficiently close to the disc outer edge (outcome Ic). Tables of individual simulation parameters and their final outcomes are given in Appendix \ref{app: simulation parameters}, and we now describe these results in more detail.

%%%%%%%%%%%%%%%%%%%%%%%%%%%%%%%%%%%%%%%%%%%%%%%%%%%%%%%%%%%%%%%%%%%%%%%%%%%%%%%%%%%%%

\subsection{Outcome Ia - the coherent disc}
\label{sec: Ia results}

\begin{figure}
  \centering
      \includegraphics[width=8cm]{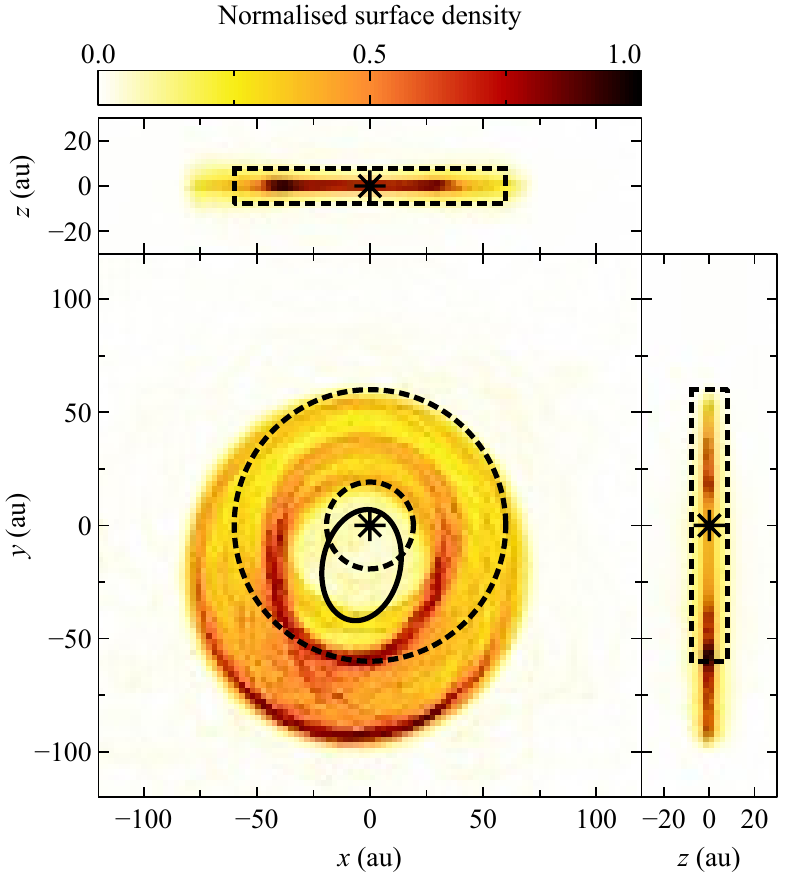}
  \caption{Instantaneous surface density of surviving disc bodies at the end of simulation 21, resulting in outcome Ia. The planet remains in the disc midplane. The black asterisk marks the central star, and the dashed and solid lines denote the initial disc edges and instantaneous orbit of the planet respectively. The orbit of each small body has been populated with $10^3$ points with randomised mean anomalies, to increase the effective number of particles on the figure. Note that care must be taken when interpreting the resonant structure (the low eccentricity ring crossing planet apocentre) plotted in this way, as addressed in Section \ref{sec: resonances}. The planet has precessed by $80^\circ$ since the start of the simulation, when its pericentre lay along the $x$ axis. The three plots have been normalised individually.}
  \label{fig: pos47}
\end{figure}

\noindent In this outcome most of the inner disc particles were ejected, and remaining material formed an elliptical disc apsidally aligned with the planet. We name this structure the ``coherent disc''. The particles themselves covered a wide range of eccentricities and orientations, and these values oscillated whilst the disc as a whole maintained its shape. This outcome was noted by \cite{Faramaz14}. The longitude of pericentre $\varpi$ (defined as $\omega + \Omega$) of both the planet and disc precessed at the same rate, and these objects remained aligned. We plot the final state of an outcome Ia system (simulation 21) with an initially coplanar planet on Figure \ref{fig: pos47}.
 
\begin{figure}
  \centering
      \includegraphics[width=6.2cm]{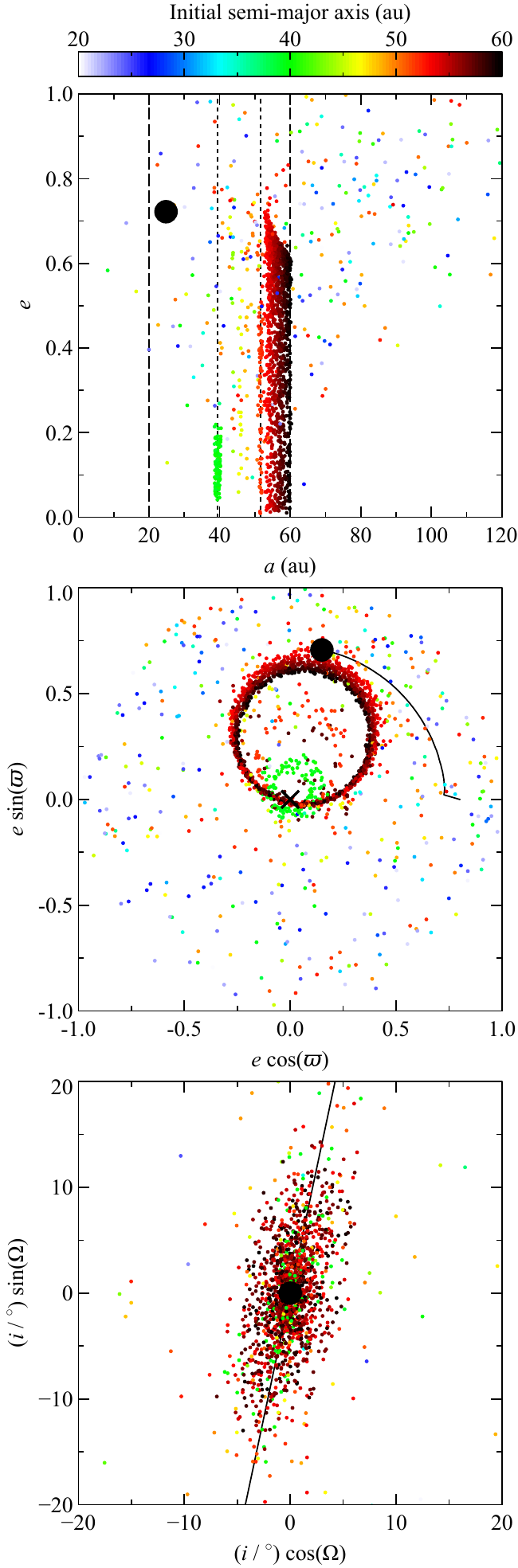}
  \caption{Instantaneous orbital elements of the particles from simulation 21. The planet's orbit is marked by the black circle, and disc particles are coloured by their initial semi-major axes. Coherent disc particles inhabit the well defined structures on these plots. Top: the dashed lines are the initial disc edges, and dotted lines are the 2:1 (innermost line) and 3:1 (outermost line) exterior resonances with the planet. Middle: the line shows the path of the planet and the cross marks the origin. Bottom: the line denotes $\Omega = \varpi_{\rm plt}$. Note that it actually intersects the planet rather than the origin. There is no obvious correlation between the position of particles on this plane and their eccentricity or longitude of pericentre. The behaviour on all plots is qualitatively similar for all eccentric, low inclination planets.}
  \label{fig: particle_elements47}
\end{figure}

We plot the final semi-major axes, eccentricities, inclinations and orientations of the simulation 21 particles on Figure \ref{fig: particle_elements47}, and see that coherent disc particles occupy a relatively well defined region in $e$ versus $a$ space. In this simulation their semi-major axes all lie between about 50 au and the initial outer disc edge at 60 au, and these did not change significantly throughout the interaction. Their eccentricities have clearly increased, and oscillate between roughly zero and some maximum ($e_{\rm max}(a) \approx 0.6-0.7$ on Figure \ref{fig: particle_elements47}) which declines with increasing semi-major axis. These imply a secular interaction, a long term angular momentum exchange leading the eccentricity oscillating whilst the semi-major axis remains constant (a detailed discription of secular theory is given in Section \ref{sec: secular_interactions}). Not all particles at smaller semi-major axes were ejected; some were long-term stable if in resonance with the planet from the start of the simulation. These particles form the additional ring structures superimposed on the coherent disc on Figure \ref{fig: pos47}.

The coherent discs are also well defined on the $e \cos{\varpi}$, $e \sin{\varpi}$ plane, implying coupling between particle eccentricity and orientation. Each moves in a circle centred on a point between the location of the planet's orbit on this plane and the origin, with those at larger semi-major axes forming tighter circles (i.e. their maximum eccentricities are smaller, as apparent from the $e$ versus $a$ plot). These circles all pass close to the origin. Each particle moves at a uniform rate anti-clockwise about its circle, and the planet and the centre of the circles also rotate in this direction (far more slowly than individual disc particles), with each circle remaining aligned with the planet. Again, this is the behaviour expected of secular particles in the low eccentricity regime (e.g \citealt{Wyatt99}). However one might not necessarily expect this behaviour if the eccentricities are large (see Section \ref{sec: secular_interactions}).

\cite{Beust14} find similar behaviour in their analysis; our particles are most akin to those below the red line on the first panel of their figure 7, which either librate or circulate in $\varpi$ whilst oscillating in eccentricity. We also note that resonant particles move in a similar manner, but with much smaller maximum eccentricities (the green points on Figure \ref{fig: particle_elements47}). Coherent disc particles also move in closed loops on the $i \cos{\Omega}$, $i \sin{\Omega}$ plane, only not in circles but ovals and in a clockwise direction. These ovals are centred on the planet and aligned with $\Omega \approx \varpi_{\rm plt}$. As $\varpi_{\rm plt}$ evolves with time, the tracks of the particles on this plane rotate. Resonant particles also behave in this manner.

The paths of secular particles on the $e \cos{\varpi}$, $e \sin{\varpi}$ and $i \cos{\Omega}$, $i \sin{\Omega}$ planes are qualitatively the same when the planet is moderately inclined to the disc midplane. For $e_{\rm plt} = 0.8$, this means planets with $i_{\rm plt} \lesssim 20$ and $25^\circ$ for $\omega_{\rm plt} = 90$ and $45^\circ$ respectively, and all inclinations for $\omega_{\rm plt} = 0^\circ$ (note that $\omega_{\rm plt} = 0^\circ$ means the planet's pericentre and apocentre both lie in the initial disc plane, whilst $\omega_{\rm plt} = 90^\circ$ means its apocentre occurs at the point on its orbit which is highest above this plane). Inclined planets cause the coherent disc to align with the planet's orbital plane, and this disc is puffed up compared to the coplanar regime. The discs also show vertical structure, appearing $x$-shaped when viewed along the planet's line of apsides. The second row of Figure \ref{fig: inclined_evolution} shows an outcome Ia interaction where the planet was initially inclined by $10^\circ$ with $\omega_{\rm plt} = 90^\circ$ (simulation 54).

Note that for very high planet inclinations with $\omega_{\rm plt} \sim 0^\circ$, particles still behave in the above manner but their inclinations may reach very high values. In this case debris forms an ovular structure enveloping the planet's orbit, but which does not have holes at the ends nor the same density pattern observed in Ib simulations. Whilst such a structure may appear similar to the Ib case, secular particle behaviour is completely akin to the Ia regime. Thus this is an extreme case of outcome Ia, with particle planes still oscillating about that of the planet, only with inclinations high enough that the ``opening angle'' of the disc (as seen front-on on the second row of Figure \ref{fig: inclined_evolution}) may be very high.

\begin{figure}
  \centering
      \includegraphics[width=8cm]{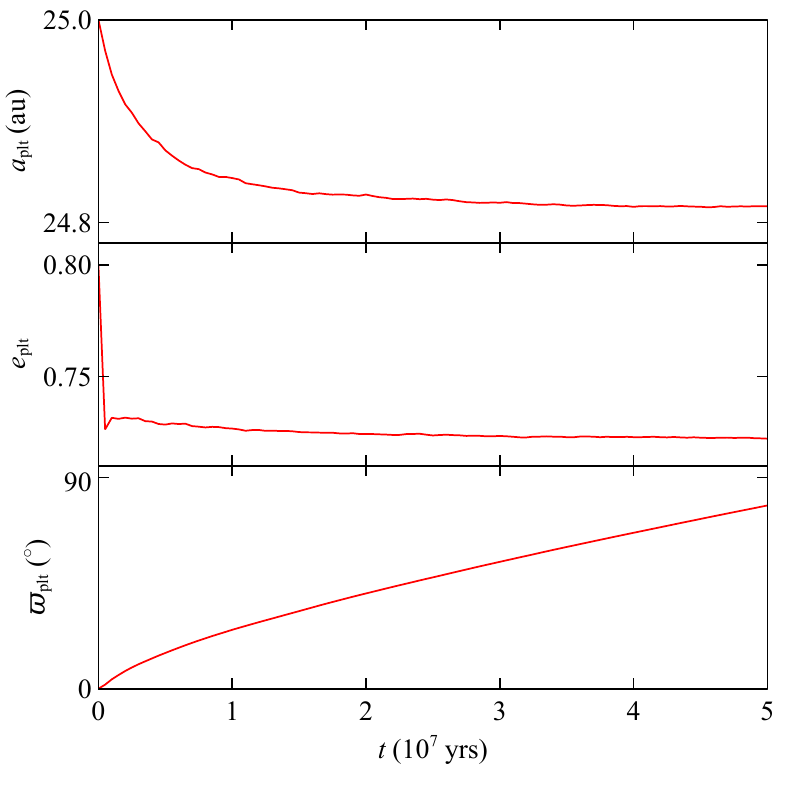}
  \caption{Orbital evolution of the planet from Figure \ref{fig: pos47} (simulation 21), which initially has $a_{\rm plt}$, $e_{\rm plt}$ and $\varpi_{\rm plt}$ equal to 25 au, 0.8 and $0^\circ$ respectively. The planet's inclination remains very close to $0^\circ$ throughout the interaction, and this evolution is also small in outcome Ia simulations featuring inclined planets.}
  \label{fig: plt_elements47}
\end{figure}

\begin{figure*}
  \centering
  \includegraphics[width=16cm]{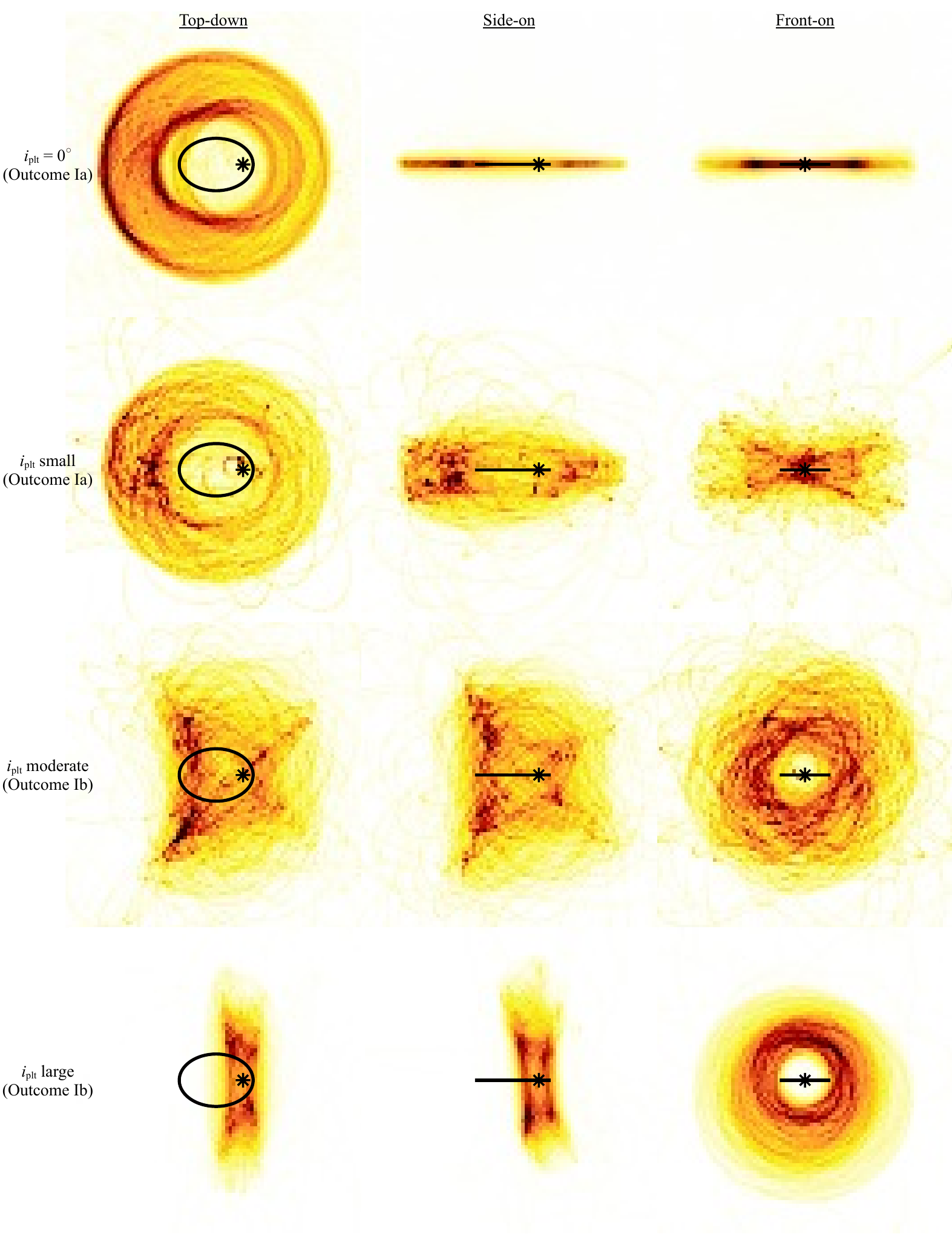}
  \caption{Dependence of debris structures on the planet's initial inclination to the disc midplane. Asterisks and solid lines denote the star and planet orbit respectively. Top: coplanar case from Figure \ref{fig: pos47} (sim. 21). Second row: same parameters, but with $i_{\rm plt} = 10^\circ$ and $\omega_{\rm plt} = 90^\circ$ (sim. 54). The disc looks similar to the coplanar case when viewed top-down but is puffed up from side-on, especially on the apocentre side. The front-on view clearly shows $x$-shaped overdensities. Third row: $i_{\rm plt} = 30^\circ$, resulting in outcome Ib (sim. 56). Debris forms a hollow bell-shaped structure with holes at both ends, looking similar if viewed top-down or side-on. Apocentre overdensity and $x$-shaped structure are observed here. Bottom row: $i_{\rm plt} = 60^\circ$, resulting in the extreme case of outcome Ib (sim. 57). The holes at either end of the bell have opened such that debris forms a disc orthogonal to the planet's orbit. All plots have been rotated so the instantaneous planet's orbit is aligned with the principle axes, have the same physical scale (the planet's major-axis is 50 au), and have been normalised individually. Each orbit is populated with $10^3$ points.}
  \label{fig: inclined_evolution}
\end{figure*}

The planet evolution is qualitatively the same for all outcome Ia interactions, and that from simulation 21 is shown on Figure \ref{fig: plt_elements47}. The semi-major axis and eccentricity typically decrease by 0.01--1 and 1--10 per cent respectively, with the latter occurring rapidly at the beginning of the interaction. The inclination change for coplanar planets is negligible. Once the planet's semi-major axis and eccentricity stop evolving its pericentre precesses at a constant rate. Inclined outcome Ia planets evolve as in the coplanar case, only now their inclinations rapidly decrease by 1-20 per cent at the start of the simulation.

%%%%%%%%%%%%%%%%%%%%%%%%%%%%%%%%%%%%%%%%%%%%%%%%%%%%%%%%%%%%%%%%%%%%%%%%%%%%%%%%%%%%%

\subsection{Outcome Ib - the bell-shaped structure}
\label{sec: Ib results}

\begin{figure*}
  \centering
      \includegraphics[width=16cm]{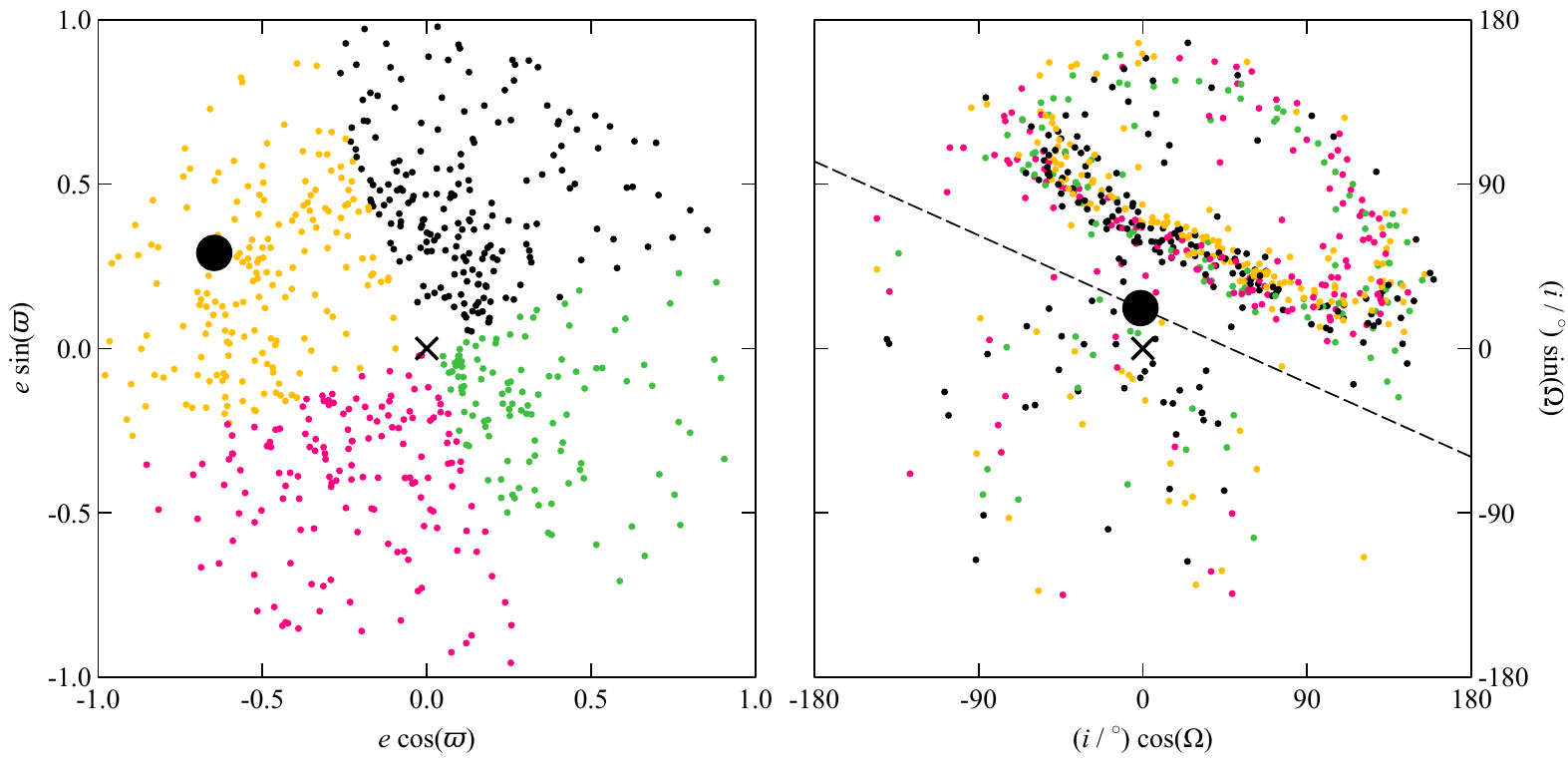}
  \caption{Instantaneous $e\cos \varpi$, $e\sin \varpi$ and $i \cos{\Omega}$, $i \sin{\Omega}$ values of particles from the outcome Ib interaction on the third row of Figure \ref{fig: inclined_evolution} (sim. 56), at 30 Myr (not the end of the simulation). Angles are defined relative to the initial disc midplane. The planet is the large black circle, and the small bodies are coloured by their pericentre orientation relative to that of the planet. Left: some roughly circular structure aligned with the planet may be discernible, with a few particles still moving about such paths. This is more apparent when viewed over multiple timesteps, and also occurs in multiple Ib simulations. However any such feature is clearly much less well defined than in the coplanar case. Right: particles forming the bell-shaped structure inhabit the squashed kidney shape. The colours show that there may be some relation between the $i \cos{\Omega}$, $i \sin{\Omega}$ values and $\varpi-\varpi_{\rm plt}$, with high inclination particles preferentially anti-aligned with the planet. The solid line is at instantaneous angle $\varpi_{\rm plt}$ to the $i \cos{\Omega}$ axis and passes through the planet's position. We empirically find that the oval remains at the same orientation with respect to such a line in all high inclination simulations (regardless of $\omega_{\rm plt}$), and both rotate about the planet's position as the planet and bell structure undergo apsidal precession.}
  \label{fig: icosO69}
\end{figure*}

\noindent If the planet's inclination is above some critical value with respect to the disc midplane (which is a function its argument of pericentre), the debris behaviour differs significantly from the previous regime. These critical inclinations were around 20 and $25^\circ$ for $\omega_{\rm plt} = 90$ and $45^\circ$ respectively in our simulations with $e_{\rm plt} = 0.8$; outcome Ib never occurred for $\omega_{\rm plt} = 0^\circ$. In outcome Ib secular particles no longer move about ovals on the $i\cos \Omega$, $i\sin \Omega$ plane, but rather a squashed ``kidney'' shape centred on a point orthogonal to the planet's orbital plane. Again the orientation of the shape is related to $\varpi_{\rm plt}$; we find it remains parallel to a line at angle $\varpi_{plt}$ to the $i\cos \Omega$ axis which passes through the planet's location, and which rotates as the planet precesses. Particle behaviour on the $e\cos \varpi$, $e\sin \varpi$ is more complicated; a number still move in circles or ellipses aligned with the planet, however most now move in much more complex shapes. Figure \ref{fig: icosO69} shows the particles from simulation 56 on both planes, at an instant (30 Myr) when structure is most clearly defined. Higher $i_{\rm plt}$ values result in tighter particle paths on the $i\cos \Omega$, $i\sin \Omega$ plane.

The physical debris structure resulting from the superposition of these particles is no longer an eccentric disc but rather a hollow bell-shaped bubble enveloping the planet's orbit, as shown on the third row of Figure \ref{fig: inclined_evolution}. The structure is broadest around planet apocentre and is roughly symmetric about its line of apsides, appearing ovular rather than $x$-shaped when viewed along this line. There are holes in both ends of the shape near planet pericentre and apocentre, and if $i_{\rm plt}$ is increased then these holes become larger and the bell structure is squashed when viewed top-down or side-on. Therefore extremely high inclinations ($i_{\rm plt} \gtrsim 60^\circ$) may result in a stable debris disc lying orthogonal to the planet's orbital plane (bottom row of Figure \ref{fig: inclined_evolution}). Particles in all outcome Ib simulations behave in the same manner on the $a$, $e$ plane as in outcome Ia, however far fewer particles are trapped in resonance in the former outcome.

The planet's eccentricity and semi-major axis undergo only minor evolution similar to that on Figure \ref{fig: plt_elements47}, with the eccentricity again decreasing rapidly at the start of the simulation. In most cases the planet's inclination decreases by less than 20 per cent, although this occurs over the timescale of the semi-major axis change (rather than the eccentricity change as for case Ia). The exceptions are simulations 57 and 58, the very highly inclined planets with stable discs orthogonal to their orbital planes. The inclinations of these planets undergo sinusoidal oscillations with periods of $\sim 20$ Myr, over which their values change by about 30 per cent. Inclined planets also eject debris at a slower rate; 80 per cent of particles were removed after 50 Myr in simulation 21, whilst simulation 56 (identical parameters but with a planet inclined by $30^\circ$) ejected only 50 per cent of disc bodies in the same time.

%%%%%%%%%%%%%%%%%%%%%%%%%%%%%%%%%%%%%%%%%%%%%%%%%%%%%%%%%%%%%%%%%%%%%%%%%%%%%%%%%%%%%

\subsection{Outcome Ic - ejection of the disc}
\label{sec: Ic results}

\noindent Some simulations result in the ejection of almost all disc particles. This outcome is favoured if the planet comes close to the outer disc edge, and is more likely for massive planets. All stable particles in outcome Ic simulations with low inclination planets lie in resonances, and occupy similar trajectories on the $e\cos \varpi$, $e\sin \varpi$ and $i \cos{\Omega}$, $i \sin{\Omega}$ planes to the resonant particles in outcome Ia. The particles do not move into such configurations, but rather remain there from the beginning of the simulations. Again few particles are caught in resonances in the inclined case, and most of those still bound at the end of outcome Ic simulations with inclined planets are in the process of being removed from the system. The planets evolve in the same manner as on Figure \ref{fig: plt_elements47}.

%%%%%%%%%%%%%%%%%%%%%%%%%%%%%%%%%%%%%%%%%%%%%%%%%%%%%%%%%%%%%%%%%%%%%%%%%%%%%%%%%%%%%
%%%%%%%%%%%%%%%%%%%%%%%%%%%%%%%%%%%%%%%%%%%%%%%%%%%%%%%%%%%%%%%%%%%%%%%%%%%%%%%%%%%%%

\section{Dynamical effects}
\label{sec: dynamical effects}

\noindent There will be three processes causing the observed behaviour: the secular (long-term) effect of the planet on the disc, the scattering of debris by the planet, and resonances between the planet and disc particles. We provide an overview of these mechanisms here, and will use them to explain our results in Section \ref{sec: analysis}.

%%%%%%%%%%%%%%%%%%%%%%%%%%%%%%%%%%%%%%%%%%%%%%%%%%%%%%%%%%%%%%%%%%%%%%%%%%%%%%%%%%%%%

\subsection{Secular interactions}
\label{sec: secular_interactions}

\noindent Secular effects cause bodies to exchange angular momentum, whilst their energies (and hence semi-major axes) remain unchanged. This mechanism may be modelled using two equivalent methods. The most common technique is to isolate secular terms in the disturbing function by ignoring those associated with the location of the bodies on their orbits. The function is then expanded in eccentricity and inclination (often to to second order, known as Laplace-Lagrange theory: see \citealt{Murray and Dermott}) or the ratio of semi-major axes. This approach is only really suited to hierarchical or low eccentricity systems where the expansion rapidly converges; for our scenario the eccentricities and semi-major axis ratios are of order unity, and hence ignoring higher order terms will be invalid. We may however use second order theory to predict behaviour to order of magnitude accuracy, and to identify how secular effects scale with different parameters.

The second approach, known as Gauss averaging, is perhaps a more intuitive description of secular perturbations. The effect of discarding terms in the disturbing function which depend on mean longitude is equivalent to extending the mass of each body around its orbit, with the density at each point weighted inversely by the body's velocity there. The system may then be thought of as several rings, and the gravitational interaction of these rings causes them to change shape and orientation. This evolution may be calculated numerically using the method of \cite{Touma09}, whereby the force of all other rings on a single point is found analytically and then these points are summed numerically over the ring. We implemented this method, using softening lengths of 1 per cent of the body's semi-major axis. This technique has the advantage that it may be applied to systems of arbitrary eccentricities and semi-major axes, however the numerical summation makes it difficult to predict outcomes without running the code. We therefore use Laplace-Lagrange theory to isolate and predict secular effects where possible, but switch to numeric Gauss averaging in cases where higher order terms significantly affect the results.

Second order theory predicts how a massless particle undergoing secular perturbations from a single planet moves on the $e \cos{\varpi}$, $e \sin{\varpi}$ plane. Its instantaneous location on this plane may be thought of as the sum of two vectors, of magnitudes $e_{\rm free}$ and $e_{\rm forced}$, at angles $\varpi_{\rm free}$ and $\varpi_{\rm forced}$ respectively to the $e \cos{\varpi}$ axis. Each particle moves at a uniform rate anti-clockwise around a circle of radius $e_{\rm free}$, the centre of which is offset by $e_{\rm forced}$ in the direction $\varpi_{\rm forced} = \varpi_{\rm plt}$. To second order in $a_{\rm plt} / a$, the forced eccentricity decreases with semi-major axis such that

\begin{equation}
e_{\rm forced}(a) \approx \frac{1}{a} A(a_{\rm plt}, e_{\rm plt}) \equiv \frac{1}{a}\left[\frac{5}{4} a_{\rm plt} e_{\rm plt}\right]
\label{eq: e_forced}
\end{equation}

\noindent (\citealt{Murray and Dermott, Mustill09}). The value of $e_{\rm free}$ is set by the initial conditions, so if a particle starts with $e=0$ then $e_{\rm free} = e_{\rm forced}$. Similar behaviour to the above is predicted for an object's inclination and longitude of ascending node, such that the particle moves in circles on the $i \cos{\Omega}$, $i \sin{\Omega}$ plane at the same rate as before, albeit in a clockwise direction.

We now compare these predictions to our simulations. The semi-major axes of coherent disc particles in outcome Ia interactions do not change significantly (typically by less than 1-5 per cent over the course of a simulation), and the same is true of the bell structure particles in outcome Ib. On the $e \cos{\varpi}$, $e \sin{\varpi}$ plane, coherent disc particles move anti-clockwise about circles so far as we can discern by eye. These circles pass close to the origin and their centres are offset by an amount scaling with $1/a$. These results agree with the predictions of low order secular theory. However the behaviour of coherent disc particles on the $i \cos{\Omega}$, $i \sin{\Omega}$ plane differs from these predictions; they move in ovals on this plane aligned with $\Omega \approx \varpi_{\rm plt}$, and hence rotate as $\varpi_{\rm plt}$ evolves (see Section \ref{sec: inclined_planets} for a description of the evolution of particle orbital planes measured relative to the orbital plane of the planet). We confirm this behaviour using the Gauss averaging method, and hence conclude that the coherent disc is a secular phenomenon. Bell-structure particles in outcome Ib also move in closed loops on the $e \cos{\varpi}$, $e \sin{\varpi}$ and $i \cos{\Omega}$, $i \sin{\Omega}$ planes, in the manner predicted by the Gauss averaging method. Hence these structures are secular phenomena too. However the shapes of these loops are much more complicated than in outcome Ia, and differ significantly from those predicted by second order theory.

\begin{figure}
  \centering
      \includegraphics[width=8cm]{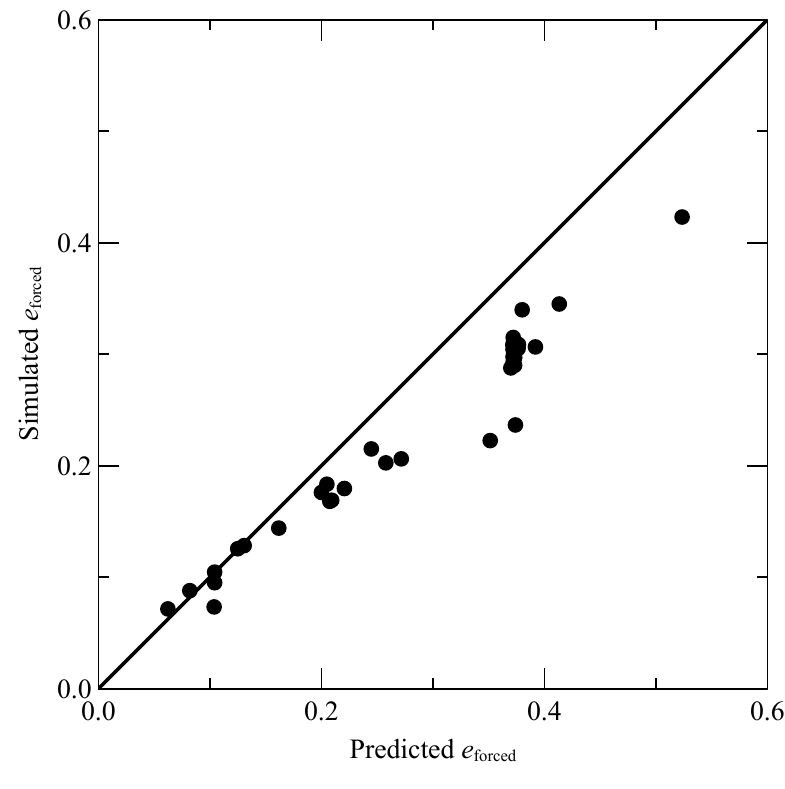}
  \caption{Forcing eccentricities of the outermost coherent disc particles in outcome Ia simulations, versus those predicted by second order secular theory (Equation \ref{eq: e_forced}). The solid line denotes a 1:1 agreement. Equation \ref{eq: e_forced} systematically overestimates $e_{\rm forced}$ at large eccentricities, however still provides a reasonable estimate if the mutual inclinations are not too high. Note that the maximum eccentricity of a particle is $\sim 2 e_{\rm forced}$. In accordance with Mustill \& Wyatt (2009), we find their equation 8 provides a worse prediction at high $e_{\rm forced}$ values than Equation \ref{eq: e_forced}.}
  \label{fig: eforced_vs_theory}
\end{figure}

We see that for outcome Ia, most predictions of second order theory appear qualitatively correct even when all bodies are significantly eccentric. On Figure \ref{fig: eforced_vs_theory} we plot the forced eccentricity of the outermost coherent disc particles from our outcome Ia simulations, versus the value predicted by Equation \ref{eq: e_forced} (the outermost particles are those least susceptible to non-secular effects). We see that the second order theory prediction is reasonably accurate even up to high forcing eccentricities; these values are within 25 per cent of the simulation values in all but three of our simulations (note that the maximum eccentricity of a particle is $e_{\rm free} + e_{\rm forced} \approx 2e_{\rm forced}$ if the disc's initial eccentricity spread $\Delta e$ is small). Thus we conclude that in the high eccentricity regime, the quantitative predictions of low order secular theory are still roughly applicable to the $a$, $e$ and $\varpi$ evolution provided mutual inclinations are not too high. However its predictions for the $i$ and $\Omega$ evolution quickly break down, so Gauss averaging must be used for these parameters. If the mutual inclination is high enough to enter the Ib regime, the quantitative predictions of second order theory fail with regard to the $e$, $\varpi$, $i$ and $\Omega$ evolution. This will be due to higher order mixed terms in the disturbing function becoming non-negligible, and significantly altering the eccentricity evolution. Therefore Gauss averaging must always be used in the Ib regime.

%%%%%%%%%%%%%%%%%%%%%%%%%%%%%%%%%%%%%%%%%%%%%%%%%%%%%%%%%%%%%%%%%%%%%%%%%%%%%%%%%%%%%

\subsection{Planet-debris scattering}
\label{sec: scattering}

\noindent Scattering exchanges both energy and angular momentum, and hence may eject bodies from the system (unlike secular interactions). Any change in a non-resonant body's semi-major axis must be a result of this effect, and we use this to find the location of the innermost stable non-resonant particles in Section \ref{sec: innermost_a}. We also use scattering arguments to constrain the planet's migration in Section \ref{sec: planet evolution}.

%%%%%%%%%%%%%%%%%%%%%%%%%%%%%%%%%%%%%%%%%%%%%%%%%%%%%%%%%%%%%%%%%%%%%%%%%%%%%%%%%%%%%

\subsection{Resonances}
\label{sec: resonances}

\noindent A particle may be long-term stable if its orbital period is a simple fraction of the planet's, hence some particles may survive strong scattering if located in resonance. The $e$ versus $a$ plot on Figure \ref{fig: particle_elements47} shows a significant number of particles trapped in the strong 2:1 external resonance, and also some in the weaker 3:1 resonance. Their behaviour on the $e \cos{\varpi}$, $e \sin{\varpi}$ plane is similar to that of the secular particles as seen on Figure \ref{fig: particle_elements47}; those in the 2:1 resonance (green) move in trajectories which resemble smaller versions of those of coherent disc particles. The 3:1 resonance particles (orange) are also visible on this plane, forming the shape aligned with the planet and located halfway between the 2:1 resonance particles and the coherent disc. Thus resonant particles form structures similar to the coherent disc in outcome Ia, which are superimposed on the latter. We observe this on Figure \ref{fig: pos47}; particles in 2:1 resonance appear to form a low eccentricity ring passing over the planet's apocentre.

\begin{figure}
  \centering
      \includegraphics[width=8cm]{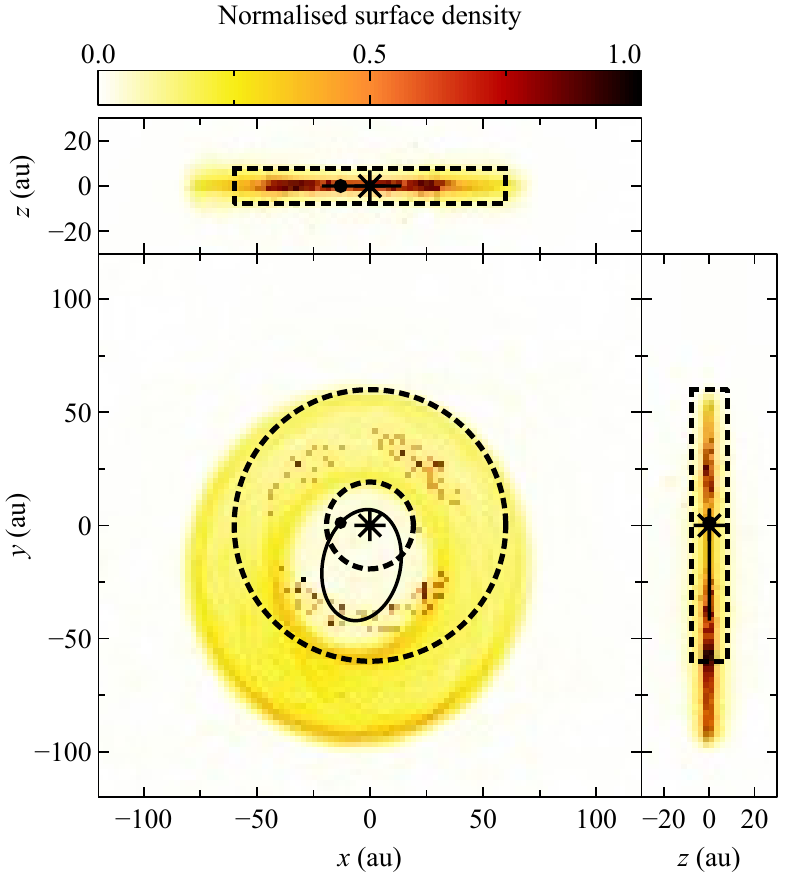}
  \caption{The 2:1 resonance structure from the outcome Ia simulation on Figure \ref{fig: pos47} (simulation 21). The instantaneous positions of 2:1 resonance particles are plotted to preserve resonant structure, and have the same weighting as the other particles. Other orbits are populated $10^3$ times as on previous figures. The asterisk, dot, solid and dashed lines denote the star, planet, planet's orbit and initial disc edges respectively.}
  \label{fig: resonant_structure}
\end{figure}

However in generating this figure we assumed all particles had random mean anomalies, which is incorrect for resonant particles as their mean anomalies are not random but depend on the instantaneous position of the planet. On Figure \ref{fig: resonant_structure} we re-plot Figure \ref{fig: pos47}, showing the instantaneous positions of particles in 2:1 resonance to preserve resonant structure. These particles were selected if their semi-major axes remained within two per cent of the nominal 2:1 resonance location for the final 0.5 Myr of the simulation. They do not form a continual ring but rather two crescents $180^\circ$ apart from one another. Resonant particles initially located in the two gaps would pass close to the planet when the latter was at apocentre, and have hence been ejected from resonance. The other particles in the 2:1 resonance never come close to the planet so are long-term stable. In addition the resonant angles of these particles librate, so the overdensities associated with material on resonant orbits oscillate between both ends of the crescent. This motion is slowest at the two extrema, hence these regions are over-dense relative to the rest of the crescent\footnote{Note that the resonant structure on Figure \ref{fig: resonant_structure} shows some similarities with the clumpy structure observed in the $\epsilon$ Eridani debris disc \citep{Greaves05}. However, more detailed characterisation of this structure is needed to determine whether there is any link.} (e.g. \citealt{Wyatt06}). Thus these particles do not actually have random mean anomalies, however we are probably justified in plotting them as a continuous ring on Figure \ref{fig: pos47} as any gaps would be difficult to observe. We have not plotted 3:1 resonance particles separately as there are fewer of these bodies and they lie farther from the star, so they do not contribute significantly to the surface density of the disc.

Owing to the small number of particles in our simulations and the thin resonant widths, resolution of all resonant structures is difficult and probably incomplete. On Figure \ref{fig: particle_elements47} seven per cent of the surviving disc particles inhabit the 2:1 resonance with a further two per cent in the 3:1, amounting to under two per cent of the initial disc mass in resonant particles. However resonant features were well resolved by \cite{Faramaz14}, who ran outcome Ia simulations but with larger numbers of particles, and we refer the reader to that paper for more detailed views of the resonant structures. We expect the coherent disc will be the dominant structure in outcome Ia systems, with less massive resonant features superimposed on top of it. Likewise particles which would otherwise be ejected may be trapped in resonance in the outcome Ib interactions, however our resolution is too low to pick out such structures on Figure \ref{fig: inclined_evolution}. In Ic systems the only surviving material would be in resonance with the planet, and so a sufficiently massive initial disc could result in a potentially observable amount of material on such orbits. 

%%%%%%%%%%%%%%%%%%%%%%%%%%%%%%%%%%%%%%%%%%%%%%%%%%%%%%%%%%%%%%%%%%%%%%%%%%%%%%%%%%%%%
%%%%%%%%%%%%%%%%%%%%%%%%%%%%%%%%%%%%%%%%%%%%%%%%%%%%%%%%%%%%%%%%%%%%%%%%%%%%%%%%%%%%%

\section{Analysis and predictions}
\label{sec: analysis}

\noindent We now characterise our results in greater detail to produce quantitative predictions for this interaction. We first examine the debris behaviour when the planet is coplanar with the disc midplane, followed by the inclined case. We then show that the evolution of the planet will almost always be small if it is at least ten times as massive as the disc, and finally we list the important timescales involved in the interaction.

%%%%%%%%%%%%%%%%%%%%%%%%%%%%%%%%%%%%%%%%%%%%%%%%%%%%%%%%%%%%%%%%%%%%%%%%%%%%%%%%%%%%%

\subsection{Coplanar planets}
\label{sec: coplanar_planets}

\noindent For coplanar planets, surviving particles are either caught in resonances or form an elliptical disc apsidally aligned with the planet. By considering the dynamical effects from Section \ref{sec: dynamical effects} we predict the location of the disc edges, the width of the unstable zone around planet apocentre (and hence differentiate between coplanar interactions resulting in outcome Ia and Ic) and characterise the non-uniform surface density of this structure.

%%%%%%%%%%%%%%%%%%%%%%%%%%%%%%%%%%%%%%%%%%%%%%%%%%%%%%%%%%%%%%%%%%%%%%%%%%%%%%%%%%%%%

\subsubsection{Disc edges}
\label{sec: disc_edges}

\begin{figure}
  \centering
      \includegraphics[width=8cm]{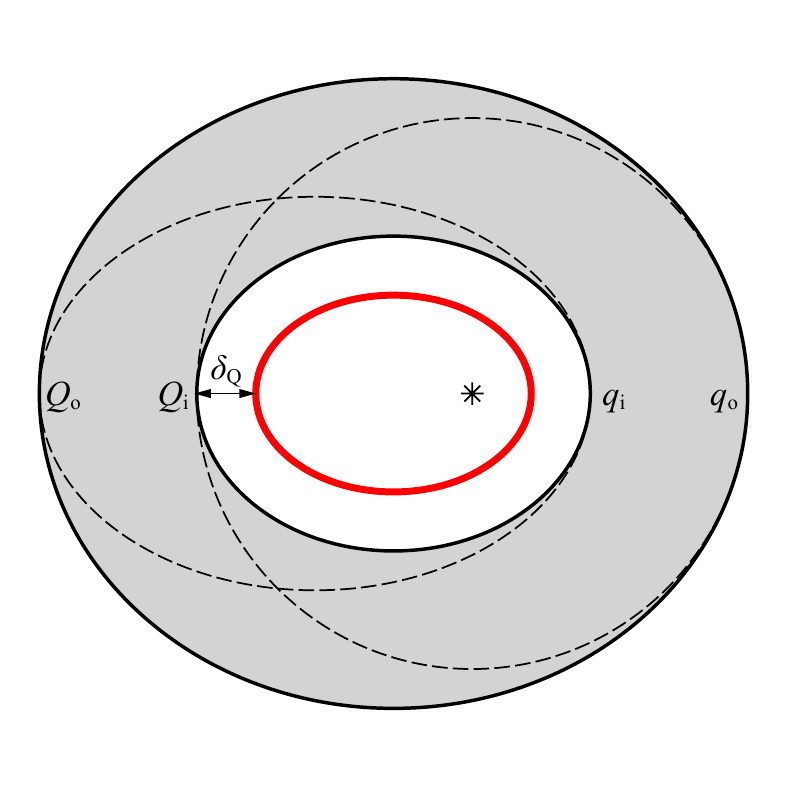}
  \caption{Schematic of the coherent disc. The star is denoted by the asterisk, and the red ellipse is the planet's orbit. The pericentre and apocentre distances of the disc's inner edge are $q_{\rm i}$ and $Q_{\rm i}$ respectively, and likewise $q_{\rm o}$ and $Q_{\rm o}$ for the outer edge. The distance between $Q_{\rm i}$ and the planet's apocentre is marked as $\delta_{\rm Q}$. The dashed lines show the two extreme orbits of the disc particles if they all had the same semi-major axis and zero eccentricity before the interaction.}
  \label{fig: disc_diagram}
\end{figure}

\noindent By considering the behaviour of debris particles, we may write down the locations of the coherent disc edges at late times. Assuming the initial eccentricity spread $\Delta e$ is small each particle's orbit evolves between a near circle and an ellipse, the latter aligned with the planet's orbit, and back again. Superimposing all intermediate orbits forms an eccentric disc, as shown on Figure \ref{fig: disc_diagram}. The edges of this disc are exactly elliptical, and we henceforth label the apocentre and pericentre of the inner and outer edges as $q_{\rm i}$, $Q_{\rm i}$, $q_{\rm o}$ and $Q_{\rm o}$ as shown on the figure. A disc formed from particles at a single semi-major axis $a$ will have edges

\vspace{-3mm}
\begin{equation}
 Q_{\rm i} = q_{\rm o} = a,
\label{eq: Qi}
\end{equation}

\vspace{-5mm}
\begin{equation}
 q_{\rm i} = a [1 - 2 e_{\rm forced}(a)],
\label{eq: qi}
\end{equation}

\noindent and

\vspace{-5mm}
\begin{equation}
 Q_{\rm o} = a [1 + 2 e_{\rm forced}(a)] .
\end{equation}

Our discs are composed of particles at various semi-major axes. If the maximum eccentricity decreases sufficiently quickly as semi-major axis increases then the inner and outer disc edges will be determined by particles with the smallest and largest semi-major axes respectively (this is satisfied by $e_{\rm forced}(a) \propto 1/a$). Extending the above equations, a coplanar coherent disc formed of particles with semi-major axes between $a_{\rm in}$ and $a_{\rm out}$ will have

\vspace{-3mm}
\begin{equation}
 Q_{\rm i} = a_{\rm in},
\label{eq: Qi extended}
\end{equation}

\vspace{-5mm}
\begin{equation}
 q_{\rm i} = a_{\rm in} [1 - 2 e_{\rm forced}(a_{\rm in})],
\label{eq: qi extended}
\end{equation}

\vspace{-5mm}
\begin{equation}
 q_{\rm o} = a_{\rm out},
\end{equation}

\noindent and

\vspace{-5mm}
\begin{equation}
 Q_{\rm o} = a_{\rm out} [1 + 2 e_{\rm forced}(a_{\rm out})]
\label{eq: Qo extended}
\end{equation}

\noindent at late times. The outermost semi-major axis will be given by $a_{\rm out} = r_2$, and we derive $a_{\rm in}$ in Section \ref{sec: innermost_a}. If the initial eccentricity spread $\Delta e$ is non-zero then $e_{\rm free}$ no longer equals $e_{\rm forced}$; the above equations are roughly modified such that $2 e_{\rm forced}$ becomes $2e_{\rm forced} + \Delta e$, $Q_{\rm i} \approx a_{\rm in}(1-\Delta e)$ and $q_{\rm o} \approx a_{\rm out}(1+\Delta e)$. Hence these are minor corrections for realistic $\Delta e$. Note that the coherent disc and planet maintain their alignment, meaning the shape of the inner disc edge is constant. Were this not the case and the precession rates differed, the planet would eventually carve out a much larger region as it precessed.

%%%%%%%%%%%%%%%%%%%%%%%%%%%%%%%%%%%%%%%%%%%%%%%%%%%%%%%%%%%%%%%%%%%%%%%%%%%%%%%%%%%%%

\subsubsection{Innermost stable particle}
\label{sec: innermost_a}

\noindent We now calculate $a_{\rm in}$, which determines the inner edge of the coherent disc, using scattering arguments. All non-resonant particles with initial semi-major axes smaller than $a_{\rm in}$ are ejected, and each particle's eccentricity is minimised when farthest from alignment with the planet (Figure \ref{fig: particle_elements47}). Therefore if the maximum eccentricity of the innermost particle is not much greater than that of the planet, these objects will be closest when the latter is at apocentre and the former is on a near circular orbit. Furthermore the planet moves slowest relative to disc particles when it is at apocentre, which increases scattering efficiency. Therefore the inner edge of the disc must be determined by scattering when the planet is at apocentre. We define a width $\delta_{\rm Q}$ such that $Q_{\rm i} \equiv Q_{\rm plt} + \delta_{\rm Q}$ (Figure \ref{fig: disc_diagram}), where $\delta_{\rm Q}$ is determined by the strength of scattering at apocentre, and we see from Equation \ref{eq: Qi extended} that the innermost stable semi-major axis also equals $Q_{\rm plt} + \delta_{\rm Q}$. Any non-resonant particles with smaller semi-major axes are unstable and will eventually be ejected from the system. Hence predicting $\delta_{\rm Q}$ would allow us to determine the innermost stable semi-major axis. We could also differentiate between outcomes Ia and Ic: if the innermost stable semi-major axis lies beyond the initial outer edge of the disc $r_2$, all non-resonant particles are unstable and will be ejected.

Instability zones around planetary orbits is a well-studied problem (e.g. \citealt{Wisdom80, Mustill12}), but most results are only valid for low planet eccentricities. We therefore establish our own criterion, based on Hill sphere arguments. We find the instantaneous Hill radius of an eccentric planet at apocentre is

\begin{equation}
R_{\rm H, Q} \approx a_{\rm plt} (1+e_{\rm plt}) \left[\frac{M_{\rm plt}}{(3-e_{\rm plt})M_*} \right]^\frac{1}{3}
\label{eq: Hill radius}
\end{equation}

\noindent (see Appendix \ref{app: hill radius derivation} for the derivation). The Hill radius at pericentre is obtained by substituting $e_{\rm plt}$ for $-e_{\rm plt}$ in the above, and it is larger at apocentre than at pericentre for all eccentricities and mass ratios. We hypothesise that the width of the instability zone is $\delta_{\rm Q} \approx n R_{\rm H, Q}$, where $n$ accounts for initially stable particles which evolve until they come close enough to the planet to be ejected. Estimating $n$ requires $\delta_{\rm Q}$ to be determined from the simulations, which is difficult to measure by eye as apparent from Figure \ref{fig: pos47}. Resonant particles and those being ejected are still present, and these must be ignored if we wish to measure the disc inner edge. There are two ways to do this: firstly we may simply state that we are not interested in particles which are either planet crossing, in strong resonance or have significant inclinations. Ignoring these leaves only the coherent disc, and $\delta_{\rm Q}$ may be ascertained from the separation of the disc inner edge and the planet's apocentre. The second method is to examine the small bodies in the $e$, $a$ plane; coherent disc particles inhabit a specific region of semi-major axis space, with a well defined maximum eccentricity and unchanged $a$ (Figure \ref{fig: particle_elements47}). Therefore omitting particles whose semi-major axes have changed by more than a few percent leaves only those in resonance and the coherent disc, hence we may read off the minimum coherent disc semi-major axis $a_{\rm in}$. We use both of these methods to calculate $\delta_{\rm Q}$ and the other edges, and both give the same answers.

\begin{figure}
  \centering
  \includegraphics[width=8cm]{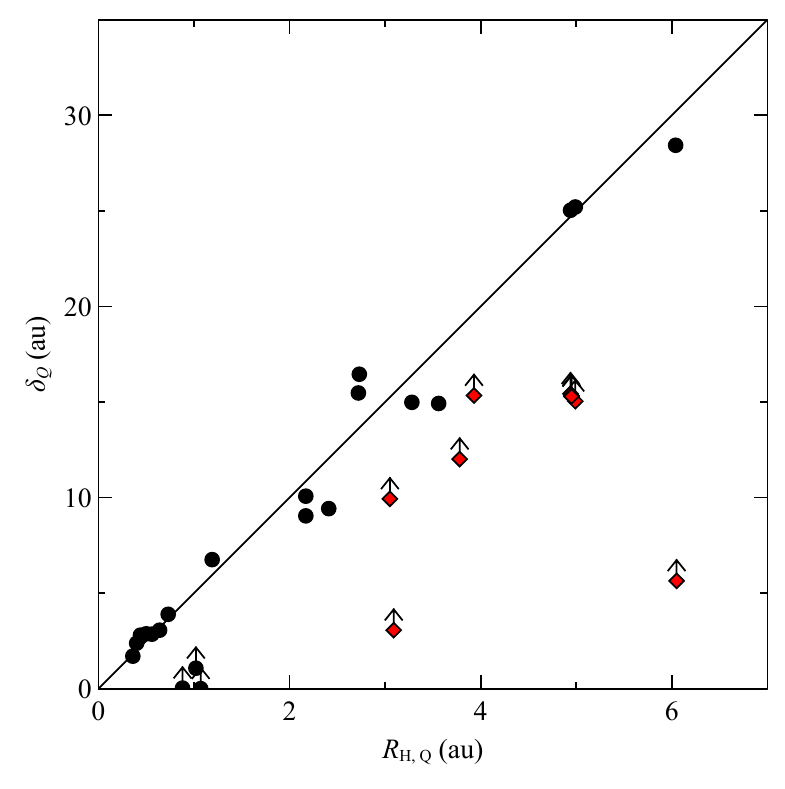}
  \caption{The distance between the planet's apocentre and the inner edge of the coherent disc $\delta_{\rm Q}$, as measured from coplanar simulations, as a function of the planet's Hill radius at apocentre (Equation \ref{eq: Hill radius}). Black circles denote outcome Ia simulations, and red diamonds show lower limits from outcome Ic simulations. The three outcome Ia lower limits, located in the bottom left of the plot, are runs where particles were still being ejected at the end of the simulation. The solid line shows $\delta_{\rm Q} = 5 R_{\rm H, Q}$.}
  \label{fig: rH_Q_scaling}
\end{figure}

We plot the values of $\delta_{\rm Q}$ versus $R_{\rm H, Q}$ for coplanar outcome Ia simulations on Figure \ref{fig: rH_Q_scaling}, as well as lower limits from coplanar outcome Ic simulations. For the latter we take the minimum value of $\delta_{\rm Q}$ to be the distance between the planet's apocentre and the initial outer edge of the disc. We find that $n \approx 5$ provides a good fit to the data, hence 

\begin{equation}
a_{\rm in} = Q_{\rm i} \approx Q_{\rm plt} + 5 R_{\rm H, Q}.
\label{eq: Qi_RH}
\end{equation}

\noindent This result is in line with several previous works on instability zones in the low eccentricity regime, which suggest that systems are stable for 10-100 Myr (the timescale of our simulations) if they are separated by of order 10 Hill radii (e.g. \citealt{Chambers96}, \citealt{Smith09}). Note that not all particles in the unstable regions may have been ejected by the end of the simulations, so the points on Figure \ref{fig: rH_Q_scaling} are more likely to be underestimations of $\delta_{\rm Q}$ than overestimations.

The cleared region will be wider at pericentre than apocentre if $2e_{\rm forced}(a_{\rm in}) < e_{\rm plt}$. Were this not the case, the innermost stable semi-major axis could potentially be determined by scattering at planet pericentre rather than apocentre. However the planet's Hill radius at pericentre is very small, so the disc would have to be very eccentric for this to occur. Therefore the inner edge of the disc will probably be determined by scattering at planet apocentre in all cases.

%%%%%%%%%%%%%%%%%%%%%%%%%%%%%%%%%%%%%%%%%%%%%%%%%%%%%%%%%%%%%%%%%%%%%%%%%%%%%%%%%%%%%

\subsubsection{Disc overdensities}
\label{sec: overdensities}

\begin{figure}
  \centering
  \includegraphics[width=8cm]{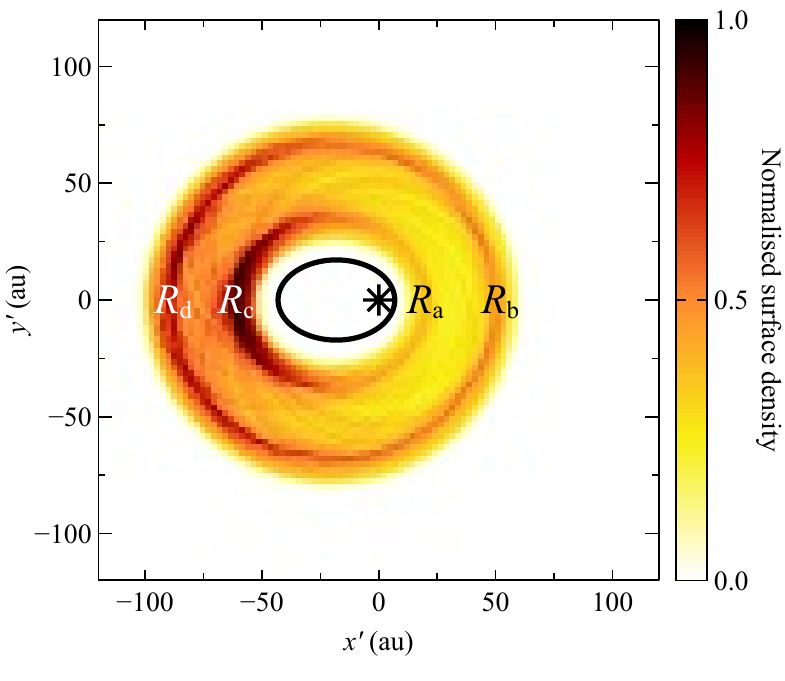}
  \caption{A coherent disc generated using the method described in Section \ref{sec: overdensities}, with $10^6$ particles and the same parameters as on Figure \ref{fig: pos47}. This reproduces the main features of the n-body coherent disc, without particles in resonance or in the process of ejection. In radial order, the locations of the overdensity peaks are denoted $R_{\rm a}$, $R_{\rm b}$, $R_{\rm c}$ and $R_{\rm d}$. The asterisk and line denote the star and planet's orbit respectively, and the coordinates are defined such that $x'$ points toward planet pericentre.}
  \label{fig: sec_plt_equiv6}
\end{figure}

\noindent We now characterise the coherent disc overdensities. To circumvent our low simulation resolution, and increase the number of coherent discs examined, we use the following method to generate these structures without running n-body simulations. Firstly we create many particles with semi-major axes between $Q_{\rm i}$ and $r_2$, following $r^{-\gamma}$ surface density profiles as before. We then generate the pre-interaction disc by assigning each particle an eccentricity between $0$ and $\Delta e$, and a randomised orientation. For a given forcing eccentricity, a particle's $e_{\rm free}$ value is the distance between its initial position on the $e \cos{\varpi}$, $e \sin{\varpi}$ plane and that of the forced eccentricity vector. Hence we find the circles on this plane about which each particle moves. Randomising particle positions around these circles assigns each an instantaneous eccentricity and orientation, and we set the inclinations to zero. Finally we randomise the mean anomaly of each particle to find its instantaneous position, and superimpose these to generate a coherent disc. We plot an example disc on Figure \ref{fig: sec_plt_equiv6}, which contains $10^6$ particles with similar parameters to those in simulation 21 (Figures \ref{fig: pos47} and \ref{fig: particle_elements47}). This shows how the disc would look at high resolution, without resonant or unstable particles.

\begin{figure*}
  \centering
  \includegraphics[width=15cm]{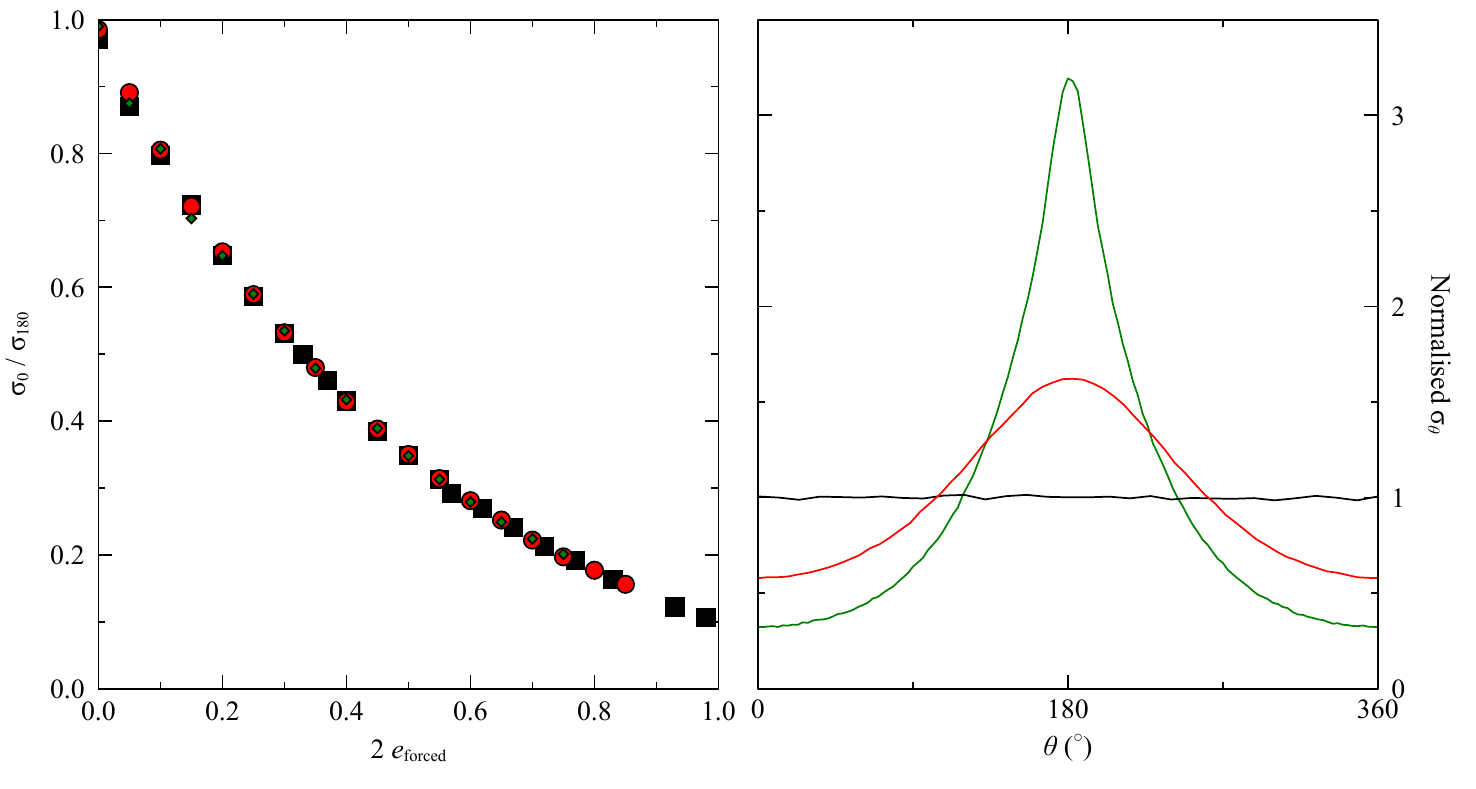}
  \caption{Left: ratio of the total cross-sectional area at pericentre and apocentre of a coplanar disc composed of particles at a single semi-major axis, as a function of forced eccentricity. The parameter $\sigma_{\theta}$ is the total cross-sectional area within a $5^\circ$ wedge radiating from the star, centred on azimuthal angle $\theta$ (degrees). Black squares, red circles and green diamonds denote discs with pre-interaction eccentricity spreads $\Delta e$ of 0, 0.1 and 0.2 respectively, and show the results to be independent of this parameter. They are also independent of semi-major axis. We find the same ratios for discs populated by particles at a range of semi-major axes, if we define the forced eccentricity to be that of a particle at the average semi-major axis. Hence this plot is applicable to any disc perturbed in a secular manner if the inclination spread is small. Right: normalised $\sigma_{\theta}$ as a function of azimuthal angle $\theta$ around the disc. Black, red and green lines denote $e_{\rm forced}$ values of 0, 0.5 and 0.98 respectively. Again, this plot is applicable to all perturbed discs with small inclination spreads.}
  \label{fig: dSD_dtheta}
\end{figure*}

The apocentre side of the coherent disc is much denser than the pericentre side, because eccentric bodies spend more time near apocentre than pericentre. We characterise this overdensity by first considering discs composed of particles at a single semi-major axis $a$. These discs are functions of $a$, $e_{\rm forced}$ and $\Delta e$, and we generated many discs using a broad range of these parameters. We then made two $5^\circ$ wedge-shaped cuts, radiating from the star and $180^\circ$ apart, through each disc. Finally we summed the particles in each wedge and divided the two, yielding the ratio of the total cross-sectional area on opposing sides of the disc. The results are shown on Figure \ref{fig: dSD_dtheta} and are independent of $a$ and $\Delta e$, only depending on forcing eccentricity. Extending this to discs composed of particles at a range of semi-major axes (which have five parameters: inner and outermost semi-major axis, $\Delta e$, $\gamma$, and $A$ such that $e_{\rm forced}(a) \equiv A / a$ from Equation \ref{eq: e_forced}) we find that the pericentre-apocentre surface density ratio is still roughly equal to that on Figure \ref{fig: dSD_dtheta}, if the forced eccentricity is now taken to be that of a particle at the mean semi-major axis. Measuring this ratio (noting that observations must be corrected for the different temperatures on both sides of the disc) would hence allow the properties of an unseen perturber to be constrained.

\begin{figure}
  \centering
  \includegraphics[width=8cm]{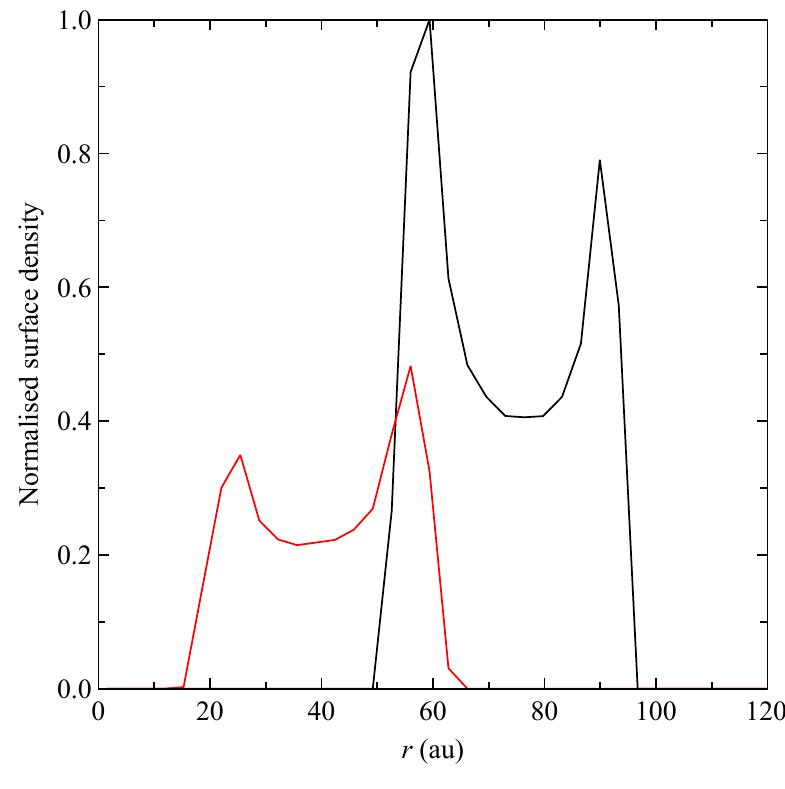}
  \caption{Surface density profile of the pericentre (red) and apocentre (black) sides of the disc on Figure \ref{fig: sec_plt_equiv6}. As on Figure \ref{fig: dSD_dtheta} the surface densities were calculated over $5^\circ$ wedges radiating from the star, centred on $\varpi = 0$ and $180^\circ$. The radial distances of the overdensity peaks $R_{\rm a}$, $R_{\rm b}$, $R_{\rm c}$ and $R_{\rm d}$ are located at 25, 55, 60 and 90 au respectively.}
  \label{fig: sec_equiv6_SD_cut}
\end{figure}

The coherent discs also exhibit finer structure. Unlike the general apocentre overdensity, these structures may not always be observable depending on the instrument beam size. However we will briefly describe these features, as they could potentially be observed with a high resolution instrument such as ALMA (e.g. \citealt{Moor13, Dent14}) or the HST (e.g. \citealt{Golimowski11}). We use the discs generated above to calculate surface density profiles over the pericentre and apocentre wedges, with an example shown on Figure \ref{fig: sec_equiv6_SD_cut}. Again, we first consider a disc formed of particles at one semi-major axis. There are either one or two overdensities on both sides of the disc, and we label their radial locations as $R_{\rm a}$, $R_{\rm b}$, $R_{\rm c}$ and $R_{\rm d}$ as on Figure \ref{fig: sec_plt_equiv6}. These peaks arise from the overlap of many orbits at these locations, where a particle's rate of eccentricity change goes from positive to negative or \textit{vice versa}. If $\Delta e = 0$ then these peaks are extremely sharp and are located at the disc edges, and if $\Delta e \neq 0$ each peak is formed from many independent peaks at slightly different locations. Hence consideration of a peak's inner and outer edges shows its width to be $2 a \Delta e$, and the same for all peaks. The two apocentre peaks will merge when the outer edge of the $R_{\rm c}$ peak overlaps the inner edge of the $R_{\rm d}$ peak, which occurs if $\Delta e \geq e_{\rm forced}$. The same result also arises on the pericentre side, so the number of peaks on both sides of the disc are equal. The ratios of peak surface densities are independent of $\Delta e$ and particle semi-major axis, but become more extreme as $e_{\rm forced}$ increases. Finally, whilst the ratios of these overdensities change their order will not; the peak with the highest surface density is always the inner peak at apocentre ($R_{\rm c}$), followed by $R_{\rm d}$, $R_{\rm b}$ and finally $R_{\rm a}$, the inner peak at pericentre, which has the lowest surface density.

Applying this analysis to discs formed of particles at multiple semi-major axes, we find each overdensity ring has a width

\vspace{-3mm}
\begin{equation}
W_{\rm peak} = \Delta a (1 + \Delta e) + 2 a_{\rm in} \Delta e.
\label{eq: peak width}
\end{equation}

\noindent Also the two overdense rings will merge if the criterion

\vspace{-3mm}
\begin{equation}
\Delta a (1 + \Delta e) + 2 a_{\rm in} \Delta e \geq 2A
\label{eq: peak  separation criterion}
\end{equation}

\noindent is satisfied, where $A$ was defined in Equation \ref{eq: e_forced}. Thus discs with a broad semi-major axis range and/or low eccentricity have a single overdensity ring, whilst a narrow semi-major axis range and/or a high eccentricity yields two rings. Hence observing the number and widths of overdensity rings in an eccentric disc would yield constraints on the perturber, in addition to those from the size and shape of the coherent disc and cleared region.

%%%%%%%%%%%%%%%%%%%%%%%%%%%%%%%%%%%%%%%%%%%%%%%%%%%%%%%%%%%%%%%%%%%%%%%%%%%%%%%%%%%%%

\subsection{Inclined planets}
\label{sec: inclined_planets}

\begin{figure}
  \centering
  \includegraphics[width=8cm]{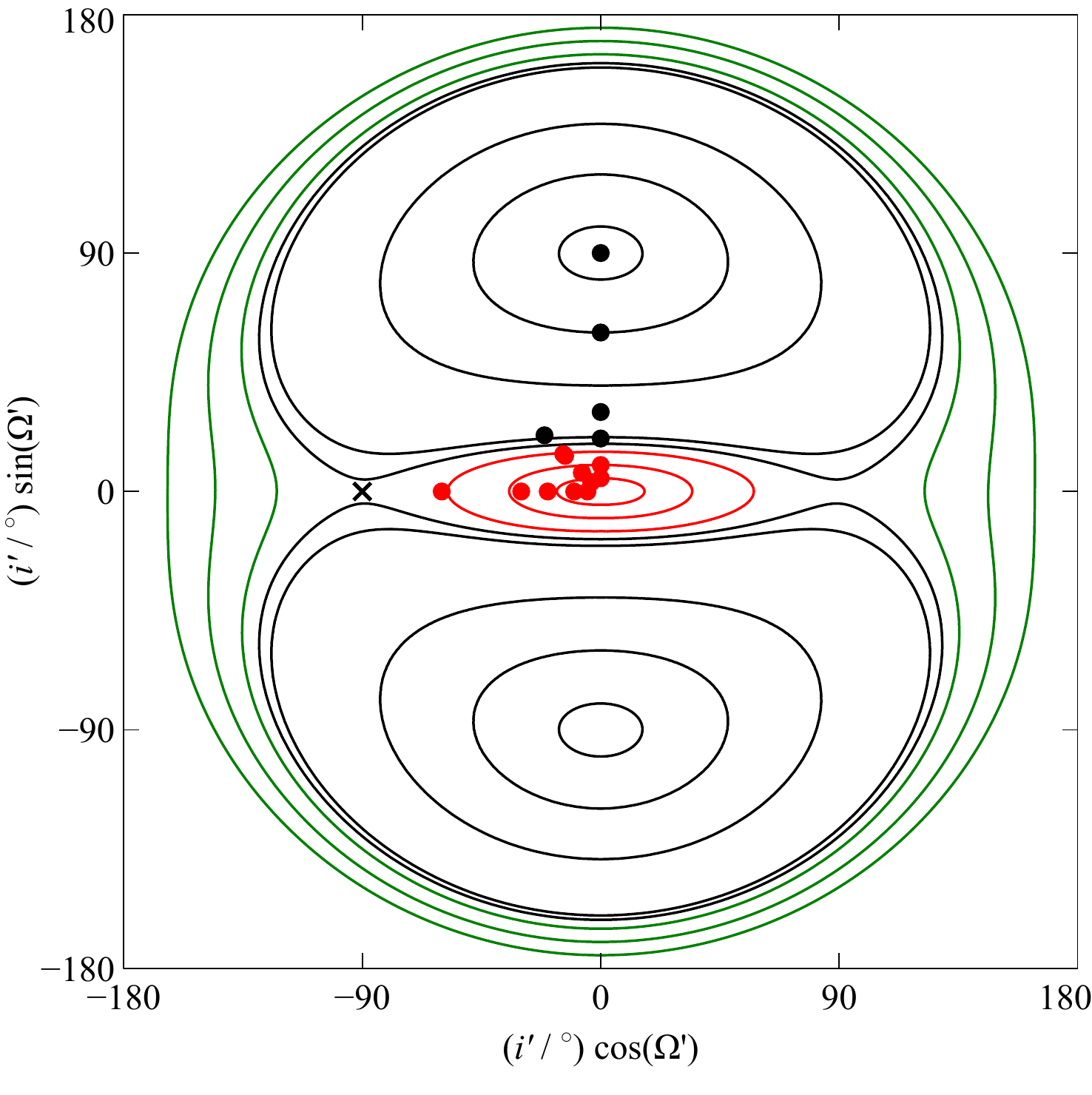}
  \caption{Secular evolution of test particle orbital planes relative to the planet's plane (with $\Omega'$ measured from the planet's pericentre direction). The lines show the paths of particles in an n-body integration where $e_{\rm plt} = 0.8$, $M_{\rm plt} / M_* = 0.0003$ and $a / a_{\rm plt} = 5$. The orbital planes of outcome Ia particles precess about that of the planet (red lines), whilst Ib particles librate about an orthogonal plane (black lines). Green lines show initially retrograde particles, a scenario not examined in this paper. Points show the initial disc midplanes in our inclined planet simulations, with the red and black circles denoting outcome Ia and Ib simulations respectively. The cross marks simulation 38, which shows Ia- and Ib-like behaviour.}
  \label{fig: iO_separatrix}
\end{figure}

\noindent Our simulations show the secular evolution of initially prograde debris follow two distinct modes, depending on the planet's initial inclination and orientation. Particle orbital planes either precess about that of the planet (outcome Ia, Figure \ref{fig: particle_elements47}), or librate about a plane orthogonal to this (Ib, Figure \ref{fig: icosO69}). These two behaviours have been noted in the literature \citep{Verrier09, Farago10, Doolin11, Kennedy12}. Figure \ref{fig: iO_separatrix} shows the secular evolution of test particle planes from an n-body simulation, where the reference (primed) frame is now the planet's orbital plane with $\Omega'$ measured from the planet's pericentre direction. For this simulation $e_{\rm plt} = 0.8$, $M_{\rm plt} / M_* = 0.0003$ and $a / a_{\rm plt} = 5$. Note that this semi-major axis ratio is larger than any from our full simulations, to isolate secular effects from those due to scattering or strong resonances. In spite of this difference the particle behaviours observed in our simulations are very similar to those on this plot, although the former show greater scatter due to the aforementioned effects.

We also plot the initial disc midplanes from our inclined simulations on Figure \ref{fig: iO_separatrix}. Because earlier we chose our reference frame to coincide with the disc's initial midplane with $x$ pointing towards the planet's initial ascending node, the orbital elements of the disc midplane relative to the planet's orbit (primed frame) are $i' = i_{\rm plt}$ and $\Omega' = 180^\circ - \omega_{\rm plt}$ or $- \omega_{\rm plt}$ for prograde and retrograde planets respectively. The conversion between these frames results in an offset and a rotation in the unprimed frame, so a particle moving along a path from Figure \ref{fig: iO_separatrix} moves about a similar shape on the $i\cos \Omega$, $i\sin \Omega$ plane but offset from the origin and orientated with the planet's line of apsides. This is exactly the behaviour observed on Figures \ref{fig: particle_elements47} and \ref{fig: icosO69}, and plotting these particles on the $i' \cos \Omega' $, $i' \sin \Omega'$ plane shows that they move along similar paths to those on Figure \ref{fig: iO_separatrix}.

Figure \ref{fig: iO_separatrix} shows that the separatrix between Ia and Ib behaviours is roughly oval shaped and orientated with its major axis along $\Omega' = 0^\circ$, so the critical planet inclination separating these behaviours is minimal for $\Omega' = \pm 90^\circ$ and maximal for $\Omega' = 0$ or $180^\circ$. In fact figure 3 from \cite{Doolin11} shows that the critical inclination is $90^\circ$ when $\Omega' = 0$ or $180^\circ$, so an initially prograde particle always lies in the Ia regime if $\omega_{\rm plt} = 0$ or $180^\circ$ (i.e. the planet's line of apsides lies in the initial disc midplane). Alternatively for $\Omega' = \pm 90^\circ$ ($\omega_{\rm plt} = \pm 90^\circ$) initially prograde particles lie in the Ib regime if the planet is sufficiently inclined; \cite{Farago10} derive this critical inclination in the $a \gg a_{\rm plt}$ limit as

\begin{equation}
i_{\rm crit}(\omega_{\rm plt} = \pm 90^\circ) = \arcsin \left(\sqrt{\frac{1-e_{\rm plt}^2}{1+4 e_{\rm plt}^2}}\right).
\label{eq: icrit}
\end{equation}

\cite{Doolin11} find the shape of test particle paths on the $i'\cos \Omega'$, $i'\sin \Omega'$ plane to be independent of semi-major axis and binary mass ratio (tested for $a/a_{\rm plt} \geq 1.5$ and $M_{\rm plt} / M_* \leq 0.5$), although not all particles are long-term stable. They also find Equation \ref{eq: icrit} is valid for all $e_{\rm plt}$, so this equation should still hold for the regime studied in this paper. Our inclined simulations use $e_{\rm plt} = 0.8$, so the above yields a critical inclination of $18.5^\circ$ for $\omega_{\rm plt} = 90^\circ$. The critical planet inclination for all other values of $\omega_{\rm plt}$ is between that of Equation \ref{eq: icrit} and $90^\circ$. All of these points agree with our results, hence we may generally predict whether outcome Ia or Ib will occur if $\omega_{\rm plt} = 0$, $\pm 90$ or $180^\circ$ (although intermediate values of $\omega_{\rm plt}$ require Figure \ref{fig: iO_separatrix} to be generated using planet specific parameters). Our only simulation which did not fall neatly into either outcome Ia, b or c is number 38, which displayed both Ia- and Ib-like behaviour. This is due to the small initial spread of particle inclinations and longitude of ascending nodes, which caused debris to split between behaviour modes as the initial midplane lay on the boundary between three behaviour regimes.

%%%%%%%%%%%%%%%%%%%%%%%%%%%%%%%%%%%%%%%%%%%%%%%%%%%%%%%%%%%%%%%%%%%%%%%%%%%%%%%%%%%%%

\subsubsection{Planets below the critical inclination}
\label{sec: low_inclination_planets}

\noindent Interactions result in outcomes Ia or Ic if the planet has a small enough initial inclination (as a function of $e_{\rm plt}$ and $\omega_{\rm plt}$) to place the system interior to the separatrix. In this case the predictions of Section \ref{sec: coplanar_planets} generally hold. Most of the surviving non-resonant particles still come closest to the planet at the latter's apocentre, so scattering at this point still sets the innermost stable semi-major axis. This value is smaller than in the coplanar case, as particles must lie closer to the star to pass within the same distance of an inclined planet and the reduced interaction time for mutually inclined bodies lowers the scattering efficiency. Hence the locations of the coherent disc edges are similar to those in the coplanar case, but the innermost semi-major axis predicted by Equation \ref{eq: Qi_RH} is now an upper limit.

A coherent disc sculpted by a low inclination planet looks very similar to the coplanar case, only now $x$-shaped overdensities are observed when viewing the disc side- or front-on (Figure \ref{fig: inclined_evolution}). These arise from the secular behaviour of the particle elements (Figure \ref{fig: particle_elements47}); particles move in ovals on the $i \cos{\Omega}$, $i \sin{\Omega}$ plane, centred on the planet and aligned with $\Omega \approx \varpi_{\rm plt}$. Thus a particle is maximally inclined to the planet when $\Omega \approx \pm \varpi_{\rm plt}$ and minimally inclined when $\Omega \approx 90^\circ \pm \varpi_{\rm plt}$. Particle orbits oscillate between these two extrema, spending more time on these orbits than on intermediate ones. Hence the superposition of particles forms an $x$ when viewed along the planet's line of apsides (front-on). This behaviour is similar to that of dust bands either side of the ecliptic in the Solar System \citep{Low84, Grogan01}. The height of the coherent disc is also greatest around planet apocentre, as a particle's eccentricity is still largest when aligned with the planet; this behaviour gives rise to overdensities when viewed from the side. Thus the overdensity analysis from Section \ref{sec: overdensities} still applies, although some additional vertical structure will also be present.

%%%%%%%%%%%%%%%%%%%%%%%%%%%%%%%%%%%%%%%%%%%%%%%%%%%%%%%%%%%%%%%%%%%%%%%%%%%%%%%%%%%%%

\subsubsection{Planets above the critical inclination}
\label{sec: high_inclination_planets}

\noindent If the prograde planet's initial inclination is at least that given by Equation \ref{eq: icrit} (depending on $\omega_{\rm plt}$), the system may lie on the other side of the separatrix and will thus result in outcome Ib or Ic. In this case the orbital planes of surviving particles librate about a point orthogonal to that of the planet, resulting in a hollow bell-shaped debris structure with holes at both ends. This outcome clearly differs from the coplanar case, however we still find several areas in which the previous results apply. Firstly the evolution of this debris remains secular in nature, evolving as predicted by the Gauss averaging code whilst particle energies stay constant. Debris still inhabits a similar region in $a$, $e$ space as in the coplanar regime, and the innermost particles are still ejected. However the divide between stable and unstable semi-major axes is now less well defined. An unstable zone still exists around the planet's orbit; a hollow inner bubble is clear of debris, and this region is roughly axisymmetric about the planet's line of apsides. The distance between the bubble edge and the planet's orbit (excluding the holes at both ends) is constant in the latter's orbital plane, and approximately five apocentre Hill radii wide - similar to the coplanar case. The distance from the star to the bell structure's outer edge is still of order the initial disc outer radius $r_2$. Finally, whilst the behaviour of particles on the $i' \cos{\Omega'}$, $i' \sin{\Omega'}$ plane differs considerably from the coplanar case, they still move in closed loops orientated with respect to the planet. Hence the parameters $i' \cos{\Omega'}$ and $i' \sin{\Omega'}$ are still strongly coupled.

This coupling of $i' \cos{\Omega'}$ and $i' \sin{\Omega'}$ gives rise to the holes at both ends of the bell; Figure \ref{fig: iO_separatrix} shows that no particles have $\Omega'$ exactly equal to 0 or $180^\circ$, hence none have nodes along the planet's line of apsides. This results in the holes, so we note that these are caused by secular rather than scattering effects. The figure also shows that the size of particle paths on the $i' \cos{\Omega'}$, $i' \sin{\Omega'}$ plane (and hence the range of $i'$ and $\Omega'$ over which particles evolve) is largest when the system only just lies inside the Ib regime. This means that the bell structure is most extended for systems where the mutual inclination between the planet's orbit and the disc midplane is $i' \approx i_{\rm crit}(\Omega')$. For systems farther inside the Ib region (towards $i' = 90^\circ$ and $\Omega' = \pm 90^\circ$) the kidney-shaped particle paths shrink, and hence the bell becomes squashed (i.e. the holes at each end become larger) owing to the increasingly narrow $i'$ and $\Omega'$ ranges over which particles oscillate. Finally in the extreme case of $i' \approx 90^\circ$ and $\Omega' \approx \pm 90^\circ$ all particles have nodes almost orthogonal to the planet's pericentre, and the bell is squashed to the extent that it becomes a disc orthogonal to the planet's orbital plane (see \citealt{Kennedy12}). Simulations 57 and 58 ($i' = 60$ and $90^\circ$ respectively, $\Omega' = 90^\circ$) have such an outcome.

Our simulations highlight the importance of a planet's argument of pericentre (i.e. the height of its pericentre above the disc midplane) in addition to its inclination in setting the secular behaviour of debris. This angle also affects the scattering of disc particles. If $\omega_{\rm plt} \approx 0^\circ$, the planet spends a lot of time near the disc midplane regardless of its inclination. Most importantly its apocentre (where scattering is strongest) lies in this plane, and so the innermost particles can be ejected within a few orbital periods. Alternatively if $\omega_{\rm plt} \approx 90^\circ$ the planet's apocentre may lie far out of the plane, and the innermost particles may survive for a few secular periods before coming close to this location. Thus the scattering efficiency is much greater for $\omega_{\rm plt} \approx 0$ or $180^\circ$ than it is for $\pm90^\circ$ (i.e. scattering is more important for $\Omega' = 0$ or $180^\circ$ than for $\Omega' =\pm 90^\circ$).

%%%%%%%%%%%%%%%%%%%%%%%%%%%%%%%%%%%%%%%%%%%%%%%%%%%%%%%%%%%%%%%%%%%%%%%%%%%%%%%%%%%%%

\subsection{Evolution of the eccentric planet}
\label{sec: planet evolution}

\noindent Having examined the final state of the debris particles, we now consider the orbital evolution of the eccentric planet. We noted that this evolution was small in our simulations, because we only examined planets at least ten times more massive than the disc. We now make a simple analytical prediction for the planet's migration distance, to show that it will be small under almost all circumstances with such mass ratios. This will justify our choice of mass ratios to study in this paper, as any larger values could potentially result in significant planetary evolution.

We use scattering arguments to constrain the planet's migration distance. As described in Section \ref{sec: scattering}, scattering is the only mechanism which may significantly change the planet's semi-major axis. The semi-major axes of surviving particles in the coplanar and highly inclined simulations are unchanged, so these objects are not being scattered. Thus if all unstable particles are eventually ejected, we may calculate the maximum change in the planet's semi-major axis using simple energy arguments. If a small mass $\delta m$ orbits a star of mass $M_*$, the energy $E$ of the system is $- G M_* \delta m / (2 a)$. Ignoring the additional small amount of energy associated with the potential of the planet on the small bodies, we find the total energy of many disc particles by summing the two body $E$ for each object. Hence for a dynamically cold disc with inner and outer radii $r_1$ and $r_2$ respectively, the total energy in disc particles between radial distances of $r_1'$ and $r_2'$ (where both radii lie within the disc) will be given by

\begin{equation}
E_{\rm r_1' < r < r_2'} = -\frac{G M_* M_{\rm disc}}{2} \frac{2-\gamma}{\gamma - 1} \frac{r_1'^{1-\gamma} - r_2'^{1-\gamma}}{r_2^{2-\gamma} - r_1^{2-\gamma}}
\label{eq: disc energy}
\end{equation}

\noindent if $\gamma \neq 1$ or 2.

\begin{figure}
  \centering
  \includegraphics[width=8cm]{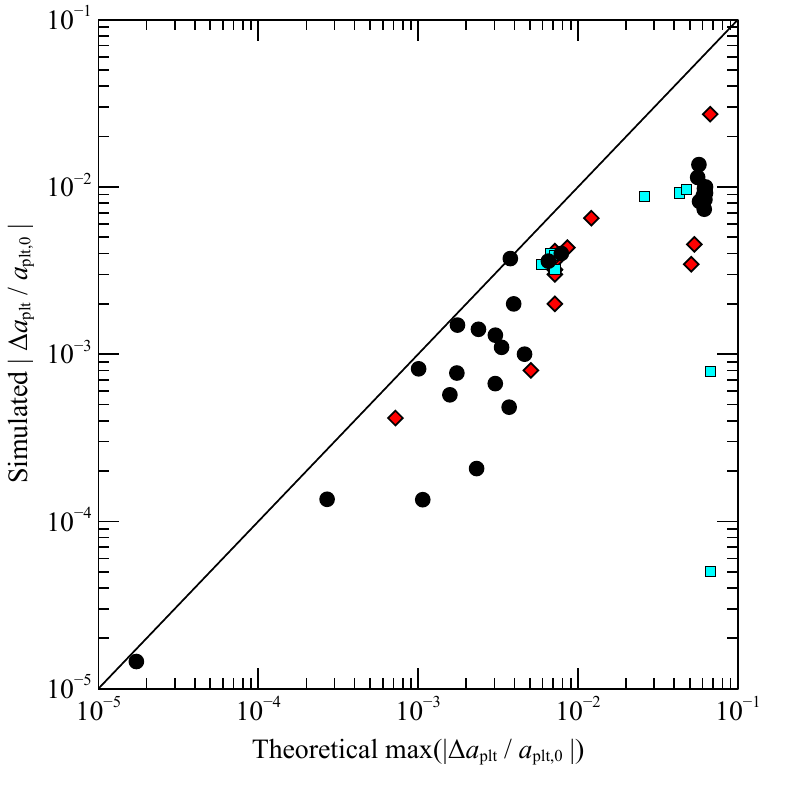}
  \caption{Maximum change in the planet's semi-major axis over each simulation, versus the upper limit from Equation \ref{eq: da_a plt}. Black circles, red squares and blue diamonds denote simulations resulting in outcomes Ia, Ib and Ic respectively, with the solid line showing a 1:1 correlation. The two squares in the lower right of the plot are simulations 57 and 58, where the planet lay almost orthogonal to the disc and caused very little scattering.}
  \label{fig: maximum migration}
\end{figure}

If all particles between $r_1'$ and $r_2'$ are \textit{just} ejected by the planet, their energy goes to zero and the planet becomes more tightly bound. Therefore equating total energies before and after scattering, we find that the planet's semi-major axis will have changed by

\begin{equation}
\frac{\Delta a_{\rm plt}}{a_{{\rm plt},0}} = \left[1 + \frac{M_{\rm disc}}{M_{\rm plt}} \Gamma \right]^{-1} - 1,
\label{eq: da_a plt}
\end{equation}

\noindent where

\begin{equation}
\Gamma \equiv a_{{\rm plt},0} \frac{2-\gamma}{\gamma - 1} \frac{r_1'^{1-\gamma} - r_2'^{1-\gamma}}{r_2^{2-\gamma} - r_1^{2-\gamma}}
\label{eq: Gamma}
\end{equation}

\noindent and $a_{{\rm plt},0}$ is the planet's initial semi-major axis. Hence if $\Gamma$ is of order of unity or lower, significant planet migration is impossible if the planet is considerably more massive than the disc. Substituting $r'_1 = r_1$ and $r'_2$ equal to the smaller of $Q_{\rm plt} + 5 R_{\rm H, Q}$ and $r_2$, we can predict the migration distances for the coplanar planets. This will be an upper limit; not all unstable particles may have been cleared by the end of the simulations, and some will remain trapped in resonances. This will also serve as an upper limit for the inclined case, where fewer bodies will have been ejected. We plot the planet migration distances against these predictions on Figure \ref{fig: maximum migration}.

By taking appropriate limits of $\Gamma$, we determine the maximum magnitude of the migration in multiple scenarios. If the disc is narrow then $0 \lesssim \Gamma \lesssim (1+e)^{-1}$, so a narrow debris ring located at planet apocentre may never lead to significant migration if the latter is more massive than the disc. For a very broad disc ($r_2 \gg r_1$), or if the planet has apocentre beyond the outer disc edge, $\Gamma \lesssim a_0 / r_1$. Therefore $\Gamma$ may be larger than unity if the inner disc edge is smaller than the planet's semi-major axis, which requires the planet to be highly eccentric. However for $a_0$ to be orders of magnitude greater than $r_1$ requires a fairly contrived set of circumstances, and we typically expect the two parameters to be of the same order. We conclude that $\Gamma$ will be of order unity or smaller in the vast majority of plausible system configurations, and hence the planet's evolution will almost always be negligible if it is at least ten times the mass of the disc.

%%%%%%%%%%%%%%%%%%%%%%%%%%%%%%%%%%%%%%%%%%%%%%%%%%%%%%%%%%%%%%%%%%%%%%%%%%%%%%%%%%%%%

\subsection{Timescales}
\label{sec: timescales}

\begin{figure*}
  \centering
  \includegraphics[width=15cm]{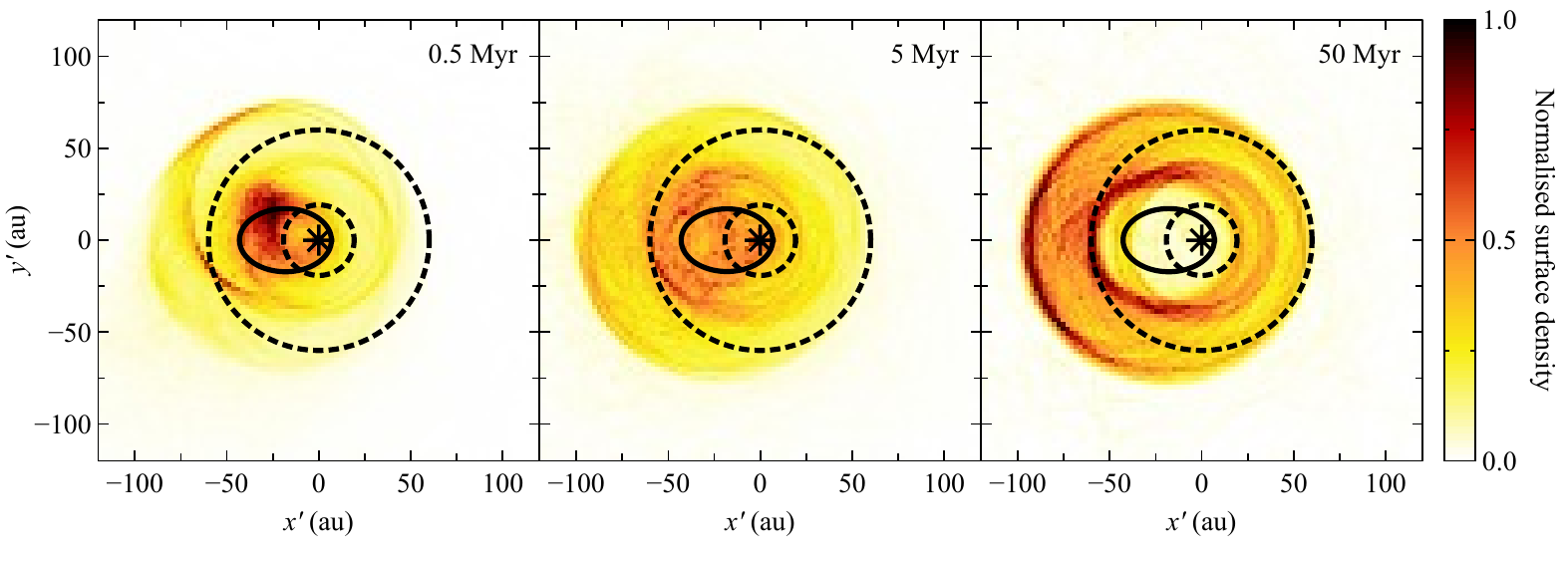}
  \caption{Evolution of the disc from Figure \ref{fig: pos47}, with the star, planet's orbit and initial disc edges shown as before. Left: at early times, spirals are clearly visible and an overdensity exists at the planet's orbit. The edges of the disc have not yet attained their final shapes, and the disc is not quite aligned with the planet. Centre: spiral structure is still just visible in the outer regions of the disc, and an overdensity still exists about the planet's orbit. Right: the disc has settled into its final configuration. These plots have been normalised individually, and the coordinates are instantaneously defined such that $x'$ points towards planet pericentre. The particle orbits have been populated as on Figure \ref{fig: pos47}. The debris clump in the first panel is caused by the secular evolution of planet-crossing particles, as described by Beust et al. (2014).}
  \label{fig: settling}
\end{figure*}

\noindent Finally we will address the important timescales in the interaction. There are several timescales of interest, and we first estimate the time taken for the coherent disc or bell structure to settle into shape. We will explain this timescale in the coplanar case for simplicity, however the conclusion is valid for the inclined planet case too. The final coherent disc shape results from the particles forming well populated circles in the $e \cos{\varpi}$, $e \sin{\varpi}$ plane. When the interaction starts the particles begin moving around the circles, at rates which depend on their semi-major axes. After some time the phase of $e$ and $\varpi$ for each particle will be roughly random, and the disc will assume its final shape. At early times however, particles at similar semi-major axes have similar phases because their secular periods only slightly differ. This means that the circles in the $e \cos{\varpi}$, $e \sin{\varpi}$ plane will not be randomly populated in phase \citep{Beust14}, and the disc will not yet have its final shape; spiral density structures may be present (\citealt{Wyatt05}, \citealt{Faramaz14}), and the inner and outer edges may not be apsidally aligned with the planet. This behaviour is visible on Figure \ref{fig: settling}, where we show the time evolution of the coherent disc from Figure \ref{fig: pos47}. The disc will settle into its final state when each particle has undergone a sufficient number of secular oscillations such that it is no longer in phase with particles at similar semi-major axes, so the timescale for this to occur will be some number (between a few and of order 10) of secular periods of the outermost disc particle. An effect of this is that the inner edge of the disc aligns with the planet quickly, whilst the outer edge takes longer to reach a stable shape. This explains why the pericentre of the outer disc occasionally differs from $r_2$ in our simulations; we note that where this is the case the simulation has ended before more than a few secular timescales of the outermost particle have elapsed.

Secondly there is the time taken for the planet to clear particles from the unstable regions. Figure \ref{fig: rH_Q_scaling} shows that particles are unstable if they come within several Hill radii of the planet's apocentre. However the secular evolution of a non-planet crossing particle periodically moves its orbit away from the planet's apocentre (see Figure \ref{fig: disc_diagram}), effectively shielding it from ejection for long periods. After a secular period the particle comes close to planet apocentre again and may be scattered, possibly repeatedly, before once again moving away through secular evolution. After one or more such interactions the particle will be ejected. Thus we expect the time taken to clear the unstable regions to be set by the longer of the secular and diffusion timescales.

Second order theory predicts a secular timescale for a test mass with semi-major axis $a$ (if $a > a_{\rm plt}$) of

\begin{equation}
t_{\rm sec} \approx 4 T_{\rm plt} \left(\frac{M_{\rm plt}}{M_*}\right)^{-1}\alpha^{-\frac{5}{2}}\left[b^{(1)}_{3/2}(\alpha)\right]^{-1},
\label{eq: secular time internal p}
\end{equation}

\noindent where $T_{\rm plt}$ is the planet's orbital period, $\alpha \equiv a_{\rm plt} / a$ and $b^{(1)}_{3/2}(\alpha)$ is a Laplace coefficient \citep{Murray and Dermott}. For the outermost unstable particle $a = a_{\rm in}$; if all non-resonant particles have been ejected in the simulation, this is taken to be $r_2$. The diffusion timescale for this particle will be of order

\begin{equation}
t_{\rm diff} \sim 0.01 T_{\rm plt} \alpha^\frac{1}{2} \left(\frac{M_{\rm plt}}{{\rm M}_*} \right)^{-2}
\label{eq: diffusion time}
\end{equation}

\begin{figure}
  \centering
  \includegraphics[width=8cm]{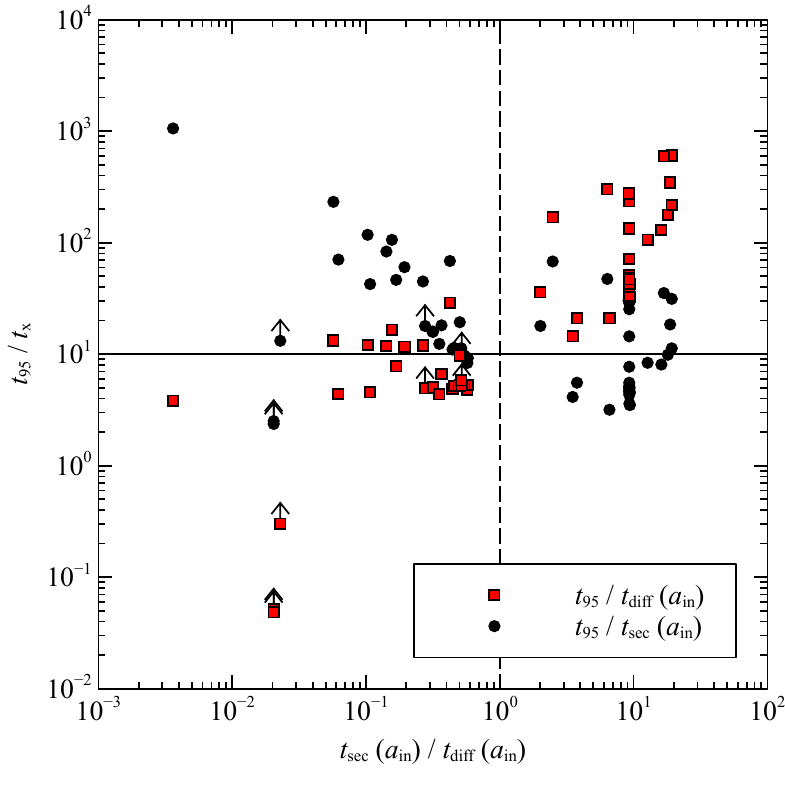}
  \caption{Time taken for 95 per cent of unstable particles to be ejected in the simulations, as functions of the secular and diffusion timescales (Equations \ref{eq: secular time internal p} and \ref{eq: diffusion time}). For simulations to the right of the dashed line the secular timescale of the outermost unstable particle is greater than its diffusion timescale, and the clearing time is roughly ten times the secular timescale. To the left of this line the clearing time is ten times the diffusion timescale. The lower limits show simulations where 95 per cent of the unstable particles are yet to be ejected.}
  \label{fig: clear_timescale}
\end{figure}

\noindent \citep{Tremaine93}. On Figure \ref{fig: clear_timescale} we plot the ratios of $t_{95}$ (the time taken to clear 95 per cent of non-resonant particles from the unstable regions) to each of these timescales, versus the ratio of the two timescales. We see that the clearing time is roughly equal to ten times the longer of the two timescales for all of our simulations. It should be noted that the majority of unstable particles are removed on timescales much shorter than this, and so the rapid decline in the planet's eccentricity at the beginning of the simulations may be attributed to the high frequency of scattering events.

Note the debris clump in the first panel of Figure \ref{fig: settling}; even though particles associated with this structure are planet crossing, their early behaviour is primarily secular in nature. Secular perturbations rapidly pump the eccentricities of these bodies towards unity, whilst their pericentres remain aligned with each other (but not with the planet). The superposition of these particles results in an elliptical overdensity misaligned with the planet's orbit. These bodies then diffuse in phase space owing to their differing secular periods as previously described, and the clump becomes symmetrical about the planet's line of apsides. This secular behaviour of planet-crossing bodies is analysed by \cite{Beust14}, and we refer the reader to their figures 6 and 7 for the physical structures formed by such particles and their corresponding phase diagrams. Eventually all non-resonant planet-crossing debris is ejected by scattering, and the inner region of the disc is cleared (final panel of Figure \ref{fig: settling}).

The third important timescale is that for system precession (see Figure \ref{fig: plt_elements47}). The coherent disc and bell structures exert secular perturbations upon the planet, causing the latter to undergo pericentre precession. However these structures are aligned with the planet. Therefore the debris causes the planet to precess, which in turn causes the debris to precess, so the whole system rotates as one. The precession period is therefore the secular period of the planet. Strictly speaking, this should be calculated using the Gauss averaging technique by summing up the perturbations from the large number of particles at different semi-major axes, eccentricities and orientations. However we find that this may be estimated to order of magnitude accuracy by modelling the debris as a single particle of semi-major axis $a_{\rm disc}$, where $Q_{\rm i} \leq a_{\rm disc} \leq r_2$, and calculating the precession rate of the planet using second order theory. For a planet perturbed by an external disc (i.e. $a_{\rm disc} > a_{\rm plt}$), the precession timescale is 

\begin{equation}
t_{\rm pre} \approx 4 T_{\rm plt} \left(\frac{f M_{\rm disc}}{M_*}\right)^{-1}\alpha^{-2}\left[b^{(1)}_{3/2}(\alpha)\right]^{-1},
\label{eq: secular time external p}
\end{equation}

\noindent where $\alpha$ is now defined as $a_{\rm plt} / a_{\rm disc}$ and $f$ is the fraction of disc particles remaining in the system. The final precession timescale predicted for simulation 21 using Equation \ref{eq: secular time external p} is $ 2\times 10^8$ years. The precession rate at the end of simulation 21 is $1.2^\circ$ / Myr (Figure \ref{fig: plt_elements47}), which implies an actual precession timescale of $ 3\times 10^8$ years. Thus the above prediction shows reasonable agreement with simulation.

%%%%%%%%%%%%%%%%%%%%%%%%%%%%%%%%%%%%%%%%%%%%%%%%%%%%%%%%%%%%%%%%%%%%%%%%%%%%%%%%%%%%%
%%%%%%%%%%%%%%%%%%%%%%%%%%%%%%%%%%%%%%%%%%%%%%%%%%%%%%%%%%%%%%%%%%%%%%%%%%%%%%%%%%%%%

\section{Discussion}
\label{sec: discussion}

\noindent We have investigated the dynamical interaction between an eccentric planet and a debris disc. In doing so we have gone further than previous studies, having examined a very broad region of parameter space and produced several quantitative results. We now discuss the \textit{caveats} and potential applications of our work.

%%%%%%%%%%%%%%%%%%%%%%%%%%%%%%%%%%%%%%%%%%%%%%%%%%%%%%%%%%%%%%%%%%%%%%%%%%%%%%%%%%%%%

\subsection{\textit{Caveats}}
 
\noindent Firstly, we did not include any gas dynamics in our simulations. We consider this to be appropriate because the interaction would likely occur after gas has dissipated from the protoplanetary disc, as gas drag would circularise forming bodies. Furthermore the removal of gas could destabilise objects, which could lead to interactions driving up a planet's eccentricity and thus act as the trigger for this scenario. 

Secondly, we only simulated a single eccentric planet. In omitting additional perturbers, we narrowed the parameter space considerably and isolated the effect of the eccentric planet on the disc and \textit{vice versa}. This also meant that we made no assumptions about how the planet became eccentric. However several scenarios would require additional bodies in the system. The most obvious mechanism to rapidly increase a planet's eccentricity is scattering by a massive second planet, most likely located at the former's pericentre. These planets would have to decouple from each other in order to prevent further scattering (also noted by \citealt{Faramaz14}), and we see that this is possible from our simulations; the eccentricity of the scattered planet decreases rapidly as it first ploughs into the disc, and an eccentricity change of $\sim 10$ per cent is possible over tens of orbital timescales (Figure \ref{fig: plt_elements47}). The much smaller semi-major axis change means the eccentric planet's pericentre rapidly moves outwards (by 2 au in the simulation on Figure \ref{fig: plt_elements47}). This could bring it far out of the scattering object's Hill radius and which could be sufficient to decouple the two bodies. One could envisage other decoupling mechanisms; for example if the scattering body moved inwards as a result of the interaction, it could then interact with interior bodies and evolve over a much shorter timescale than the more distant eccentric planet. In fact there need not be additional objects at all; if planetary mergers occur, the result may be a single body on a highly eccentric orbit (e.g. \citealt{Lin97}). Thus the results of our investigation should be broadly correct for many cases even if additional planets are present.

However, if there were additional perturbers in the system which remain coupled with the eccentric planet, these could repeatedly scatter that planet. The latter's orbit would therefore change with each interaction. This would probably prevent the coherent disc or bell structure from forming, as particles would rapidly switch between stability and instability. Many more particles (and possibly the eccentric planet) would be ejected in this regime. Long-term stable particles would also lie farther from the planets, so the resulting debris structure would probably not be as extreme as in the single planet case.

Even in the absence of further planet-planet scattering, secular and resonant interactions in a multi-planet system could result in the outermost eccentric planet (which is presumably that considered here) continually evolving, and hence the system may not have a well defined final state as in our simulations. Even without planet evolution, the debris behaviour could still differ significantly in this regime. For example, secular perturbations would not necessarily cause particles to be apsidally aligned with the outermost planet, since the longitude of forced pericentre $\varpi_{\rm forced}$ (which determines the orientations of the centres of the circular particle tracks on the $e \cos{\varpi}$, $e \sin{\varpi}$ plane) would depend on the secular solution for all of the planetary bodies. In the coplanar case this would cause the degree of alignment between the eccentric planet and coherent disc to vary, which would result in a larger innermost stable semi-major axis. Three (or more) body resonances could also trap particles in different locations to the two body resonances in the single planet case, and the reduced secular times could lead to more rapid clearing and disc settling if more than one planet were present.

Finally our simulations only model the largest debris particles, for which gravity is the significant force. Smaller dust grains are also affected by radiation pressure, which is not included here. Optical images only show these small grains, hence the observed debris structures may differ in appearance to those seen in our simulations. As such our results are more comparable to longer wavelength (i.e. ALMA) images of larger particles, and act as tracers of the parent body dynamics.

%%%%%%%%%%%%%%%%%%%%%%%%%%%%%%%%%%%%%%%%%%%%%%%%%%%%%%%%%%%%%%%%%%%%%%%%%%%%%%%%%%%%%

\subsection{Applications}

\noindent We showed that material surviving an encounter with an eccentric planet forms structures which would not otherwise be expected, and our work could potentially predict the properties and dynamical history of such unseen planets from observed debris. Highly eccentric, coplanar planets force surviving material into a coherently eccentric disc, which may be significantly elliptical. Observing such a structure could be indicative of an unseen eccentric planet in roughly the same plane as the disc and aligned with it. A sharp inner disc edge, possibly combined with resonant structure, would additionally suggest that the planet is strongly scattering disc material; this could be evidence of previous planetary evolution, which would bring the planet close enough to the disc to remove non-resonant material and maintain a sharp disc edge. In the opposite case of a highly inclined eccentric planet, the resulting bell-shaped debris structure is even more unusual. Observing this structure would not only yield the orientation of the planet's current orbit (debris is symmetric above and below the planet's orbital plane, and is offset from the star in the direction of planet apocentre) but could also point towards a violent dynamical history; this structure only forms if the planet is significantly inclined to the disc at early times, which could hint at some major planetary evolution in the system's past. Our results could also identify an ongoing planet-disc interaction; a debris structure with a global offset, spiral structure and a central, stationary overdensity could suggest a planet has recently been placed onto an eccentric disc crossing orbit (Figure \ref{fig: settling}).

Beyond the simple application above, our numerical results could also place more detailed constraints on a sculpting planet if debris has settled into its final state. If the disc and planet lie in the same plane, the forcing eccentricity of a particle forming the inner disc edge may be calculated from the edge shape using Equations \ref{eq: Qi extended} and \ref{eq: qi extended} (assuming the pre-interaction disc was dynamically cold). Using either second order secular theory (Equation \ref{eq: e_forced}) or Gauss averaging would then constrain the planet's eccentricity as a function of its semi-major axis, and assuming the planet is no longer disc crossing yields further constraints. The planet must be massive enough to have cleared the inner regions within the age of the system, yielding two lower mass limits as functions of semi-major axis (Equations \ref{eq: secular time internal p} and \ref{eq: diffusion time}). Finally the width of the instability region (Equations \ref{eq: Hill radius} and \ref{eq: Qi_RH}) gives another mass constraint. Thus we may bound the mass, semi-major axis, eccentricity, orientation and orbital plane of an interior perturbing planet using the shape of the coherent disc alone. Resonant structures superimposed on this disc would yield additional semi-major axis constraints, and could even point to the planet's current location on its orbit. Finally density variations within the disc could produce additional constraints on forcing eccentricity (Figure \ref{fig: dSD_dtheta} and Equations \ref{eq: peak width} and \ref{eq: peak  separation criterion}). Similar constraints may be found even if the planet is initially inclined with respect to the disc, as many of the numeric results still hold in this regime (see Sections \ref{sec: low_inclination_planets} and \ref{sec: high_inclination_planets}). Note however that Gauss averaging must be used in the high inclination regime, as the quantitative predictions of second order secular theory fail here. 

\begin{figure*}
  \centering
  \includegraphics[width=15cm]{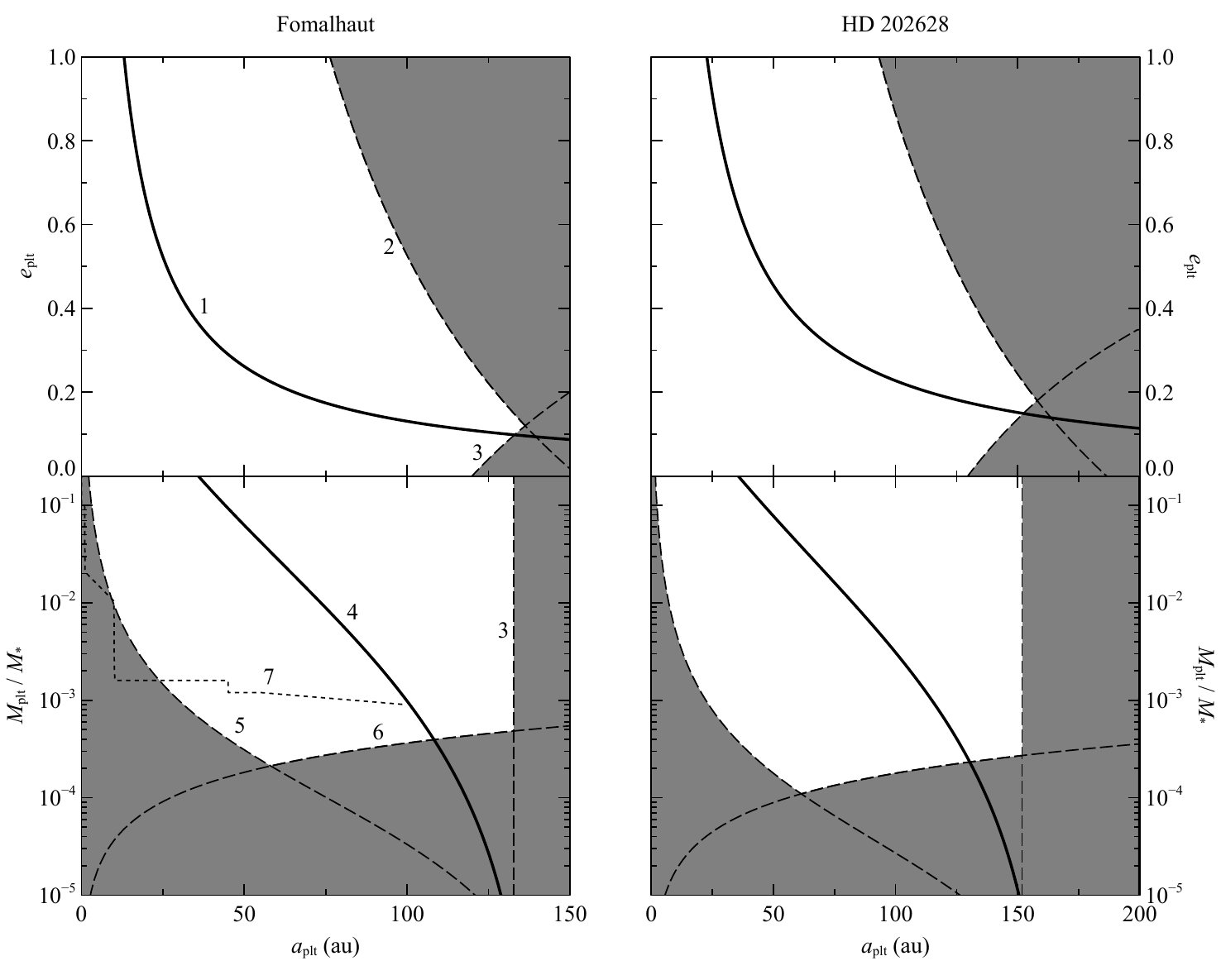}
  \caption{Constraints on a perturbing planet from the observed discs of Fomalhaut (left) and HD 202628 (right), were they to be explained via our results. The planets are expected to lie along the solid lines, where line 1 results from the forced eccentricity of the disc (Equation \ref{eq: e_forced}) and line 4 shows the constraint $Q_i \approx Q_{\rm plt} + 5 R_{\rm H, Q}$ (Equation \ref{eq: Qi_RH}). Grey regions are excluded, with lines 2 and 3 showing the limits for the planet to lie interior to the disc inner edge. Lines 5 and 6 denote $10 t_{\rm sec} < t_{\rm obs}$ and $10 t_{\rm diff} < t_{\rm obs}$ respectively, where $t_{\rm obs}$ is the system age. The dotted line (7) is the observational upper mass limit for a planet orbiting Fomalhaut \citep{Kenworthy13}. Solutions with $e_{\rm plt} \geq 1$ are unphysical. We do not not claim to explain these systems, but we use them as examples of how our results could potentially be applied.}
  \label{fig: example_systems}
\end{figure*}

Large numbers of debris discs have now been imaged, with many displaying significant asymmetries and/or stellar offsets. We apply the above method to the HD 202628 and Fomalhaut systems as examples (we stress that we are not actually attempting to explain these systems via this scenario, but rather to show how our results could potentially be applied). Both systems host resolved, eccentric debris rings with sharp, elliptical inner edges. For our purposes we assume both rings to be remains of initially broader, circular discs truncated by hypothetical eccentric planets with apocentres interior to their inner edges. The lack of significant vertical structure implies the hypothetical planets are coplanar (whilst such discs could also occur if $i_{\rm plt} \approx \omega_{\rm plt} \approx 90^\circ$, coplanar planets seem much more likely). Starting with Fomalhaut, \cite{Kalas13} finds that the deprojected inner disc edge may be fitted with an ellipse of semi-major axis 138 au and eccentricity 0.12. Assuming a 1.92 ${\rm M}_\odot$ star of age 450 Myr \citep{Mamajek12} we plot the constraints derived using the above method on Figure \ref{fig: example_systems}, along with the observational upper mass limits for the hypothetical object \citep{Kenworthy13}. We find that such an object must have a semi-major axis of $100-110$ au, an eccentricity of $0.10 - 0.13$ and a mass of $0.77 - 1.73$ Jupiter masses ($M_{\rm Jupiter}$) to sculpt the observed disc inner edge. These predictions compare favourably with those of \cite{Chiang09}. However problems arise as we consider this system in more detail; we have neither explained the highly eccentric Fomalhaut-b nor included the additional perturbations it could exert on debris \citep{Tamayo14,Beust14}. Observed dust also forms a narrow ring whilst our model predicts a broad disc, which we also cannot explain; it is implicit in our model that the proper eccentricities are comparable with the forced eccentricities, which is not the case for Fomalhaut. Therefore we can probably rule out our model as an explanation for the Fomalhaut system, for which it would be more appropriate to study using system-specific simulations.

Figure \ref{fig: example_systems} also shows our constraints on an eccentric perturber in the HD 202628 system, derived for an elliptical inner disc edge of semi-major axis 158 au and eccentricity 0.18, with a solar mass star of age 2.3 Gyr \citep{Krist12}. We found no upper mass limits in the literature, but if we assume the perturber is a planet rather than a brown dwarf (i.e. $M_{\rm plt} \lesssim 10 M_{\rm Jupiter}$) then the hypothetical object must have a semi-major axis of $80-130$ au, an eccentricity of $0.17-0.28$ and a mass $> 0.2 M_{\rm Jupiter}$. Thus our model could potentially explain this system. Whilst really more suited to larger disc eccentricities than observed in either of the above systems, these examples show that our results may be used to quickly estimate the major parameters of a hypothetical perturber using just a few observables. These could then form the basis of more detailed system specific studies.

Discs detected in the far-infrared and sub-millimetre (such as $\zeta^2$ Reticuli: \citealt{Eiroa10}) often exhibit what could be interpreted as more extreme asymmetries to those seen in Fomalhaut and HD 202628, and more akin to those predicted here. However, there remains some uncertainty as to whether the offset emission in such systems arises from circumstellar material or from background galaxies (e.g. \citealt{Panic13}). Galaxies have a similar temperature to that expected for cold dust at $\sim 100$ au from a star, and are ubiquitous enough to be commonly found in close proximity to nearby stars. Thus it is often difficult to distinguish between these two interpretations (e.g. \citealt{Nilsson10}). Our study shows is that it is not possible to rule out a circumstellar origin for the emission by virtue of its non-axisymmetry about the star, since we find that highly asymmetric disks are a natural outcome in systems with planets on highly eccentric orbits. Given the large number of highly eccentric extrasolar planets detected, and the significant eccentricity excitation predicted by Solar System evolution models and some Hot Jupiter formation theories (e.g. \citealt{Boley12}), highly eccentric discs could be more common than one might naively expect. The abundances of both eccentric planets and misshapen debris discs will help determine the frequency of systems with eccentric planets encountering discs, and our work provides testable predictions for such a scenario.

%%%%%%%%%%%%%%%%%%%%%%%%%%%%%%%%%%%%%%%%%%%%%%%%%%%%%%%%%%%%%%%%%%%%%%%%%%%%%%%%%%%%%
%%%%%%%%%%%%%%%%%%%%%%%%%%%%%%%%%%%%%%%%%%%%%%%%%%%%%%%%%%%%%%%%%%%%%%%%%%%%%%%%%%%%%

\section{Conclusions}
\label{sec: conclusions}

\noindent We used n-body simulations and analytics to characterise the interaction between an eccentric planet and a less massive debris disc. We identify a chaotic zone of predictable width around the planet's orbit, and any non-resonant debris here will be ejected within $\sim 10$ secular or diffusion times (whichever is longer). Surviving non-resonant debris undergoes secular evolution, forming either an eccentric disc, a hollow bell-shaped structure or a polar disc, depending on whether the planet's initial inclination with respect to the disc is low, moderate or high respectively. These structures are apsidally aligned with the planet and centred on its orbital plane, and form within a few secular times. They also contain overdensities at predictable locations, which could be used to constrain parameters of an unseen perturber. The planet's dynamical evolution will almost always be negligible in this mass regime. Our results may be used to predict the general outcomes of this scenario, or to infer the properties of unseen eccentric planets from observed debris features.

%%%%%%%%%%%%%%%%%%%%%%%%%%%%%%%%%%%%%%%%%%%%%%%%%%%%%%%%%%%%%%%%%%%%%%%%%%%%%%%%%%%%%
%%%%%%%%%%%%%%%%%%%%%%%%%%%%%%%%%%%%%%%%%%%%%%%%%%%%%%%%%%%%%%%%%%%%%%%%%%%%%%%%%%%%%

\section{Acknowledgements}
\noindent We thank Dan Tamayo for his insightful review, which was a great help in improving this paper. TDP acknowledges the support of an STFC studentship, and MCW is grateful for support from the European Union through ERC grant number 279973.

%%%%%%%%%%%%%%%%%%%%%%%%%%%%%%%%%%%%%%%%%%%%%%%%%%%%%%%%%%%%%%%%%%%%%%%%%%%%%%%%%%%%%
%%%%%%%%%%%%%%%%%%%%%%%%%%%%%%%%%%%%%%%%%%%%%%%%%%%%%%%%%%%%%%%%%%%%%%%%%%%%%%%%%%%%%

%%%%%%%%%%%%%%%%%%%%%%%%%%%%%%%%%%%%%%%%%%%%%%%%%%%%%%%%%%%%%%%%%%%%%%%%%%%%%%%%%%%%%
%%%%%%%%%%%%%%%%%%%%%%%%%%%%%%%%%%%%%%%%%%%%%%%%%%%%%%%%%%%%%%%%%%%%%%%%%%%%%%%%%%%%%

\appendix

\section{Simulation parameters}
\label{app: simulation parameters}

\noindent Tables \ref{tab: coplanar simulations} and \ref{tab: inclined simulations} list the initial conditions and outcomes of our n-body simulations.

\begin{center}
\begin{table*}
\begin{tabular}{c c c c c c c c c}

\hline
Simulation & $a_{\rm plt}$ (au) & $e_{\rm plt}$ & $r_1$ (au) & $r_2$ (au) & $M_{\rm plt}$ (${\rm M}_\oplus$) & $M_{\rm disc}$ (${\rm M}_\oplus$) & Outcome & Notes \\
\hline
1	&	25	&	0.8	&	20	&	60	&	10	&	1	&	Ia	&		\\
2	&	25	&	0.8	&	20	&	60	&	10	&	0.1	&	Ia	&		\\
3	&	25	&	0.8	&	20	&	60	&	100	&	10	&	Ia	&		\\
4	&	25	&	0.8	&	20	&	60	&	1000	&	10	&	Ic	&		\\
5	&	25	&	0.8	&	20	&	60	&	1000	&	10	&	Ic	&	Ended at 8 Myr	\\
6	&	25	&	0.8	&	20	&	120	&	1000	&	20	&	Ia	&		\\
7	&	15	&	0.67	&	20	&	60	&	1000	&	10	&	Ia	&		\\
8	&	15	&	0.67	&	20	&	120	&	1000	&	10	&	Ia	&		\\
9	&	25	&	0.8	&	20	&	200	&	1000	&	1	&	Ia	&		\\
10	&	25	&	0.8	&	45	&	120	&	1000	&	10	&	Ia	&		\\
11	&	25	&	0.8	&	20	&	60	&	1000	&	10	&	Ic	&	$M_* = 2{\rm M}_\odot$	\\
12	&	25	&	0.8	&	20	&	60	&	1000	&	1	&	Ic	&		\\
13	&	25	&	0.8	&	40	&	60	&	1000	&	10	&	Ic	&		\\
14	&	30	&	0.83	&	20	&	60	&	1000	&	10	&	Ic	&		\\
15	&	17.5	&	0.71	&	20	&	60	&	1000	&	10	&	Ia	&		\\
16	&	25	&	0.2	&	20	&	60	&	1574	&	10	&	Ia	&	Ended at 5 Myr	\\
17	&	29.6	&	0.83	&	54.1	&	139.2	&	1000	&	0.1	&	Ia	&		\\
18	&	25.8	&	0.81	&	5	&	96.1	&	100	&	0.1	&	Ia	&		\\
19	&	14.3	&	0.65	&	8.3	&	170.3	&	100	&	1	&	Ia	&		\\
20	&	18	&	0.72	&	19.8	&	31.6	&	1000	&	100	&	Ic	&		\\
21	&	25	&	0.8	&	20	&	60	&	100	&	10	&	Ia	&	$10^4$ particles	\\
22	&	25	&	0.8	&	20	&	60	&	1000	&	10	&	Ic	&	$10^4$ particles	\\
22	&	40	&	0.9	&	65	&	85	&	100	&	10	&	Ic	&	Ended at 500 Myr	\\
23	&	35	&	0.8	&	55	&	70	&	100	&	10	&	Ic	&	Ended at 500 Myr	\\
24	&	20	&	0.9	&	35	&	39.5	&	10	&	1	&	Ia	&	Ended at 500 Myr	\\
25	&	42.5	&	0.88	&	20	&	60	&	1000	&	10	&	Ic	&		\\
26	&	6.3	&	0.2	&	7.5	&	25	&	100	&	1	&	Ia	&		\\
27	&	6.7	&	0.25	&	7.5	&	25	&	100	&	1	&	Ia	&		\\
28	&	7.1	&	0.3	&	7.5	&	25	&	100	&	1	&	Ia	&		\\
29	&	7.7	&	0.35	&	7.5	&	25	&	100	&	1	&	Ia	&		\\
30	&	8.3	&	0.4	&	7.5	&	25	&	100	&	1	&	Ia	&		\\
31	&	9.1	&	0.45	&	10	&	25	&	100	&	1	&	Ia	&		\\
32	&	10	&	0.5	&	10	&	25	&	100	&	1	&	Ia	&		\\
\hline

\end{tabular}

\caption{Initial conditions for simulations where the planet was initially coplanar with the disc midplane. $r_1$ and $r_2$ denote the initial inner and outer disc radii respectively. Unless otherwise stated the simulations had solar mass stars, $10^3$ particles and ended after 50 Myr. The maximum inclination of the disc particles was $5^\circ$ in all simulations. Outcome Ia denotes the case where the innermost non-resonant particles were cleared and remaining debris formed an eccentric disc apsidally aligned with the planet, and Ic is where all non-resonant particles were ejected or in the process of being ejected by the end of the simulation.}

\label{tab: coplanar simulations}
\end{table*}
\end{center}

\begin{center}
\begin{table*}
\begin{tabular}{c c c c c c c c c c}

\hline
Simulation & $a_{\rm plt}$ (au) & $e_{\rm plt}$ & $i_{\rm plt}$ ($^\circ$) & $\omega_{\rm plt}$ ($^\circ$) & $r_1$ (au) & $r_2$ (au) & $M_{\rm plt}$ (${\rm M}_\oplus$) & $M_{\rm disc}$ (${\rm M}_\oplus$) & Outcome \\
\hline
33	&	25	&	0.8	&	5	&	0	&	20	&	60	&	100	&	10	&	Ia	\\
34	&	25	&	0.8	&	10	&	0	&	20	&	60	&	100	&	10	&	Ia	\\
35	&	25	&	0.8	&	20	&	0	&	20	&	60	&	100	&	10	&	Ia	\\
36	&	25	&	0.8	&	30	&	0	&	20	&	60	&	100	&	10	&	Ia	\\
37	&	25	&	0.8	&	60	&	0	&	20	&	60	&	100	&	10	&	Ia	\\
38	&	25	&	0.8	&	90	&	0	&	20	&	60	&	100	&	10	&	-	\\
39	&	25	&	0.8	&	5	&	0	&	20	&	60	&	1000	&	10	&	Ic	\\
40	&	25	&	0.8	&	10	&	0	&	20	&	60	&	1000	&	10	&	Ic	\\
41	&	25	&	0.8	&	19	&	0	&	20	&	60	&	1000	&	10	&	Ic	\\
42	&	25	&	0.8	&	30	&	0	&	20	&	60	&	1000	&	10	&	Ic	\\
43	&	25	&	0.8	&	30	&	0	&	20	&	70	&	1000	&	10	&	Ia	\\
44	&	25	&	0.8	&	5	&	45	&	20	&	60	&	100	&	10	&	Ia	\\
45	&	25	&	0.8	&	10	&	45	&	20	&	60	&	100	&	10	&	Ia	\\
46	&	25	&	0.8	&	20	&	45	&	20	&	60	&	100	&	10	&	Ia	\\
47	&	25	&	0.8	&	30	&	45	&	20	&	60	&	100	&	10	&	Ib	\\
48	&	25	&	0.8	&	5	&	45	&	20	&	60	&	1000	&	10	&	Ic	\\
49	&	25	&	0.8	&	10	&	45	&	20	&	60	&	1000	&	10	&	Ic	\\
50	&	25	&	0.8	&	19	&	45	&	20	&	60	&	1000	&	10	&	Ic	\\
51	&	25	&	0.8	&	30	&	45	&	20	&	60	&	1000	&	10	&	Ib/Ic	\\
52	&	25	&	0.8	&	30	&	45	&	20	&	70	&	1000	&	10	&	Ib	\\
53	&	25	&	0.8	&	5	&	90	&	20	&	60	&	100	&	10	&	Ia	\\
54	&	25	&	0.8	&	10	&	90	&	20	&	60	&	100	&	10	&	Ia	\\
55	&	25	&	0.8	&	20	&	90	&	20	&	60	&	100	&	10	&	Ib	\\
56	&	25	&	0.8	&	30	&	90	&	20	&	60	&	100	&	10	&	Ib	\\
57	&	25	&	0.8	&	60	&	90	&	20	&	60	&	100	&	10	&	Ib	\\
58	&	25	&	0.8	&	90	&	90	&	20	&	60	&	100	&	10	&	Ib	\\
59	&	25	&	0.8	&	5	&	90	&	20	&	60	&	1000	&	10	&	Ic	\\
60	&	25	&	0.8	&	10	&	90	&	20	&	60	&	1000	&	10	&	Ic	\\
61	&	25	&	0.8	&	19	&	90	&	20	&	60	&	1000	&	10	&	Ib	\\
62	&	25	&	0.8	&	30	&	90	&	20	&	60	&	1000	&	10	&	Ib	\\

\hline

\end{tabular}

\caption{Initial conditions of the simulations with inclined planets. All had solar mass stars, $10^3$ particles, maximum initial particle inclinations of $5^\circ$ with respect to the disc midplane and finished after 50 Myr. The initial values of $\Omega_{\rm plt}$ and $\omega_{\rm plt}$ were both $90^\circ$. Outcome Ib results in the bell-shaped structure shown on the last two rows of Figure \ref{fig: inclined_evolution}. Outcomes Ia and Ic are described in Table \ref{tab: coplanar simulations}. The outcome of simulation 38 is described in Section \ref{sec: high_inclination_planets}.}
\label{tab: inclined simulations}
\end{table*}
\end{center}

%%%%%%%%%%%%%%%%%%%%%%%%%%%%%%%%%%%%%%%%%%%%%%%%%%%%%%%%%%%%%%%%%%%%%%%%%%%%%%%%%%%%%

\section{Derivation of the eccentric Hill radius}
\label{app: hill radius derivation}

\begin{figure}
  \centering
  \includegraphics[width=8cm]{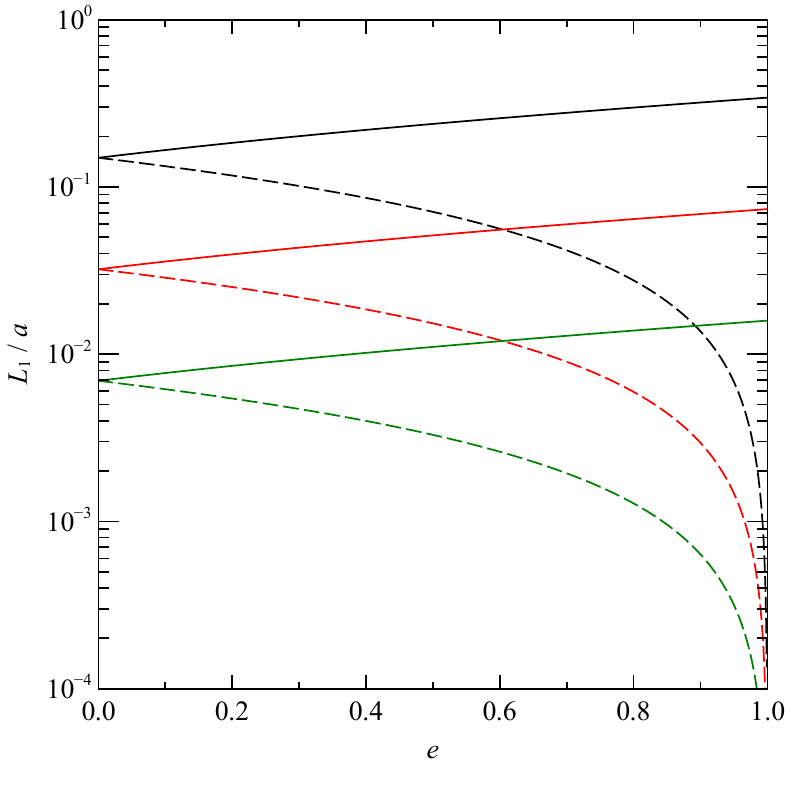}
  \caption{Distance from an eccentric secondary mass to the first Lagrange point at pericentre (dashed lines) and apocentre (solid lines), taken to be the instantaneous Hill radius. The black, red and green lines denote secondary to primary mass ratios of $10^{-2}$, $10^{-4}$ and $10^{-6}$ respectively.}
  \label{fig: L1 points}
\end{figure}

\noindent For circular orbits the Hill radius may be derived by determining the distance to the first Lagrange point, where the combined gravitational force of the planet and star on a test mass, as well as any fictitious forces arising from system rotation, exactly balance. However the standard concept of the Hill radius does not work in the eccentric case, because the planet's changing radial distance and velocity means that the width of this region varies over the orbit. Nonetheless we may still calculate the ``instantaneous'' Hill radius of an eccentric planet at a point on its orbit using the same arguments. A similar approach was taken by \cite{Hamilton91,Hamilton92} to find the instantaneous Hill radius at pericentre, only they ommited velocity terms which affect the final result. Here we find the instantaneous Hill radii at pericentre and apocentre without omitting such terms.

Care must be taken with our definitions of the instantaneous Lagrange points for an eccentric body; in the circular case, these were derived by balancing the gravitational and rotational forces. For the eccentric case however we note that these forces should \textit{not} exactly balance, because the planet is accelerating inwards at apocentre and outwards at pericentre. Therefore the radial acceleration of the test mass at pericentre and apocentre should match that of the planet if the former is at the first Lagrange point. Noting that the centrifugal force is the only non-zero rotational force at these points (see \citealt{Hamilton91,Hamilton92} for more details), we arrive at

\begin{equation}
\frac{G M_1}{R_1^2} - \frac{G M_2}{L_1^2} - \Phi^2 R = \ddot{r},
\label{eq: L1 force balance}
\end{equation}

\noindent where $G$ is the gravitational constant, $M_1$ and $M_2$ are the primary (star) and secondary (planet) masses respectively, $L_1$ is the distance between the secondary and the test mass (the distance to the first Lagrange point), $R_1$ is the distance between the test mass and the primary, $\Phi$ is the instantaneous angular velocity of the system, $R$ is the distance between the test mass and the system centre of mass, and $r$ is the instantaneous distance between the primary and secondary. We may calculate $\ddot{r}$ by considering the radial acceleration of the secondary, yielding

\begin{equation}
\ddot{r} = \frac{G (M_1 + M_2)}{r^2} - \Phi^2 r.
\label{eq: ddot r}
\end{equation}

\noindent Note that this is positive at pericentre and negative at apocentre. The system angular velocity at apocentre will be

\begin{equation}
\Phi^2 = \frac{G (M_1 + M_2)}{a^3 (1+e)^3}(1 - e),
\label{eq: Phi}
\end{equation}

\noindent where $a$ and $e$ are the semi-major axis and eccentricity of the $M_1$ - $M_2$ binary. The angular velocity at pericentre may be obtained by substituting $e$ for $-e$ in the above equation. Finally it is clear that $R_1$ and $R$ are given by $r - L_1$ and $M_1 / (M_1+M_2) r - L_1$ respectively.

Substituting these into Equation \ref{eq: L1 force balance}, and making the approximation that $L_1 \ll r$ as $M_2 / M_1 \ll 1$, we arrive at the following cubic for the first Lagrange point at apocentre

\begin{equation}
z^3 \left[3-e+\frac{M_2}{M_1}(1-e)\right] - z^2 \left[\frac{M_2}{M_1}e\right] - \frac{M_2}{M_1} \approx 0,
\label{eq: Hill cubic}
\end{equation}

\noindent where $z \equiv L_1 / r$ and is small. Again, the equation for the pericentre case is obtained by substituting $e$ for $-e$ in the above. The $z^2$ term is negligible and may be discarded, hence we find that the Hill radius at apocentre is given by

\begin{equation}
R_{\rm H, Q} \approx a(1+e) \left[\frac{M_2}{(3-e)M_1} \right]^\frac{1}{3},
\label{eq: Hill radius appendix}
\end{equation}

\noindent which reduces to the circular case if $e=0$. The Hill radius at pericentre is obtained by substituting $e$ for $-e$, and we show the Hill radius at pericentre and apocentre on Figure \ref{fig: L1 points}. If we repeat this analysis for the second Lagrange point (which lies along the primary - secondary axis but on the far side of the planet), we find that both of these points are approximately equidistant from the smaller mass, as in the circular case.

\label{lastpage}

\end{document}